\pdfoutput=1
\documentclass[journal, 10pt]{IEEEtran}

\usepackage{amsmath}
\usepackage{fixmath}
\usepackage{amsfonts}

\usepackage{enumerate,letltxmacro}
\LetLtxMacro\itemold\item
\usepackage{amssymb}
\usepackage{setspace}

\usepackage{float}
\usepackage{subfig}
\usepackage{pbox}

\usepackage[table, rgb]{xcolor}
\usepackage{tikz, pgfplots}
\usepackage{verbatim}
\usepackage{cite,url,bm}
\usepackage{tablefootnote, threeparttable, booktabs, multirow}
\usepackage{algorithmicx}
\usepackage[ruled]{algorithm}
\usepackage{algpseudocode}

\alglanguage{pseudocode}
\usepackage{psfig}
\usepackage{latexsym}

\usepackage{graphicx}
\usepackage{tikz}

\usetikzlibrary{arrows,decorations.pathmorphing,backgrounds,positioning,fit,petri}

\usetikzlibrary{shapes,arrows}

\usepackage[english]{babel}
\pgfplotsset{footnotesize}
\usepackage{parskip}
\usepackage{pgf}
\usepackage[normalem]{ulem}
\usepackage[autoplay]{animate}

\usepackage{lipsum}

\usepackage{dutchcal}

\DeclareFontFamily{OT1}{pzc}{}
\DeclareFontShape{OT1}{pzc}{m}{it}{<-> s * [1] pzcmi7t}{}
\DeclareMathAlphabet{\mathpzc}{OT1}{pzc}{m}{it}

\usepackage[mathscr]{eucal}
\DeclareMathAlphabet\mathbfcal{OMS}{cmsy}{b}{n}

\def\bmt{\left[\begin{matrix}}

\def\emt{\end{matrix}\right]}
\def\bx{\mathbcal{x}}
\def\by{\mathbcal{y}}
\def\bg{\mathbcal{g}}

\def\bb{\mathbf{b}}

\def\bu{\mathbcal{u}}

\def\bz{\mathbcal{w}}
\def\and{\text{~and~}}

\def\etal{\textit{et al.}}
\def\R{\mathbb{R}}

\def\bA{\mathbf{A}}

\def\bA{\mathbfcal{A}}
\def\bD{\mathbfcal{D}}
\def\bM{\mathbfcal{M}}
\def\bN{\mathbfcal{N}}
\def\bP{\mathbfcal{P}}
\def\bX{\mathbfcal{X}}
\def\bY{\mathbfcal{Y}}

\def\tF{\mathcal{S}}
\def\tF{g}

\def\bxt{\mathbcal{x}^{(t)}}
\def\byt{\mathbcal{y}^{(t)}}
\def\bDt{\mathbfcal{D}^{(t)}}

\makeatletter
\newsavebox\myboxA
\newsavebox\myboxB
\newlength\mylenA
\newcommand*\lbar[2][.75]{%
    \sbox{\myboxA}{$\m@th#2$}%
    \setbox\myboxB\null
    \ht\myboxB=\ht\myboxA%
    \dp\myboxB=\dp\myboxA%
    \wd\myboxB=#1\wd\myboxA
    \sbox\myboxB{$\m@th\overline{\copy\myboxB}$}
    \setlength\mylenA{\the\wd\myboxA}
    \addtolength\mylenA{-\the\wd\myboxB}%
    \ifdim\wd\myboxB<\wd\myboxA%
       \rlap{\hskip 0.5\mylenA\usebox\myboxB}{\usebox\myboxA}%
    \else
        \hskip -0.3\mylenA\rlap{\usebox\myboxA}{\hskip 0.3\mylenA\usebox\myboxB}%
    \fi}
\makeatother

\def\R{\mathbb{R}}
\def\bY{\mathbf{Y}}
\def\vec{\text{vec}}

\usepackage{fancyhdr}

\begin{document}
\pagestyle{headings}
%





\title{Classifying Multi-channel UWB SAR Imagery via Tensor
Sparsity Learning Techniques}
\author{Tiep Vu,~\IEEEmembership{Student Member,~IEEE,}  Lam
Nguyen,~\IEEEmembership{Member,~IEEE,} Vishal
Monga,~\IEEEmembership{Senior Member,~IEEE} 
}
\maketitle


\begin{abstract} 

Using low-frequency (UHF to L-band) ultra-wideband (UWB) synthetic aperture radar (SAR) technology for detecting buried and obscured targets, e.g. bomb or mine, has been successfully demonstrated recently. Despite promising recent progress, a significant open challenge is to distinguish obscured targets from other (natural and manmade) clutter sources in the scene. The problem becomes exacerbated in the presence of noisy responses from rough ground surfaces. In this paper, we present three novel sparsity-driven techniques, which not only exploit the subtle features of raw captured data but also take advantage of the polarization diversity and the aspect angle dependence information from multi-channel SAR data. First, the traditional sparse representation-based classification (SRC) is generalized to exploit shared information of classes and various sparsity structures of tensor coefficients for multi-channel data. Corresponding tensor dictionary learning models are consequently proposed to enhance classification accuracy. Lastly, a new tensor sparsity model is proposed to model responses from multiple consecutive looks of objects, which is a unique characteristic of the dataset we consider. Extensive experimental results on a high-fidelity electromagnetic simulated dataset and radar data collected from the U.S. Army Research Laboratory side-looking SAR demonstrate the advantages of proposed tensor sparsity models.
 
\end{abstract}

\IEEEpeerreviewmaketitle

\section{Introduction}
\label{sec:intro}

Over the past two decades, the U.S. Army has been investigating the capability
of low-frequency, ultra-wideband (UWB) synthetic aperture radar (SAR) systems
for the detection of buried and obscured targets in various applications, such
as foliage penetration~\cite{lam1997}, ground penetration~\cite{lam1998}, and
sensing-through-the-wall~\cite{lam2008}. These systems must operate in the
low-frequency spectrum spanning from UHF frequency band to L band to achieve
both resolution and penetration capability. Although a lot of progress has been
made over the years, one critical challenge that
low-frequency UWB SAR technology still faces is discrimination of targets of
interest from other natural and manmade clutter objects in the scene. The key
issue is that that the targets of interest are typically small compared to the
wavelength of the radar signals in this frequency band and have very low radar
cross sections (RCSs). Thus, it is very difficult to discriminate targets and
clutter objects using low-resolution SAR imagery.

\subsection{SAR geometry and image formation overview}

\begin{figure}[t]
\centering
\includegraphics[width = .495\textwidth]{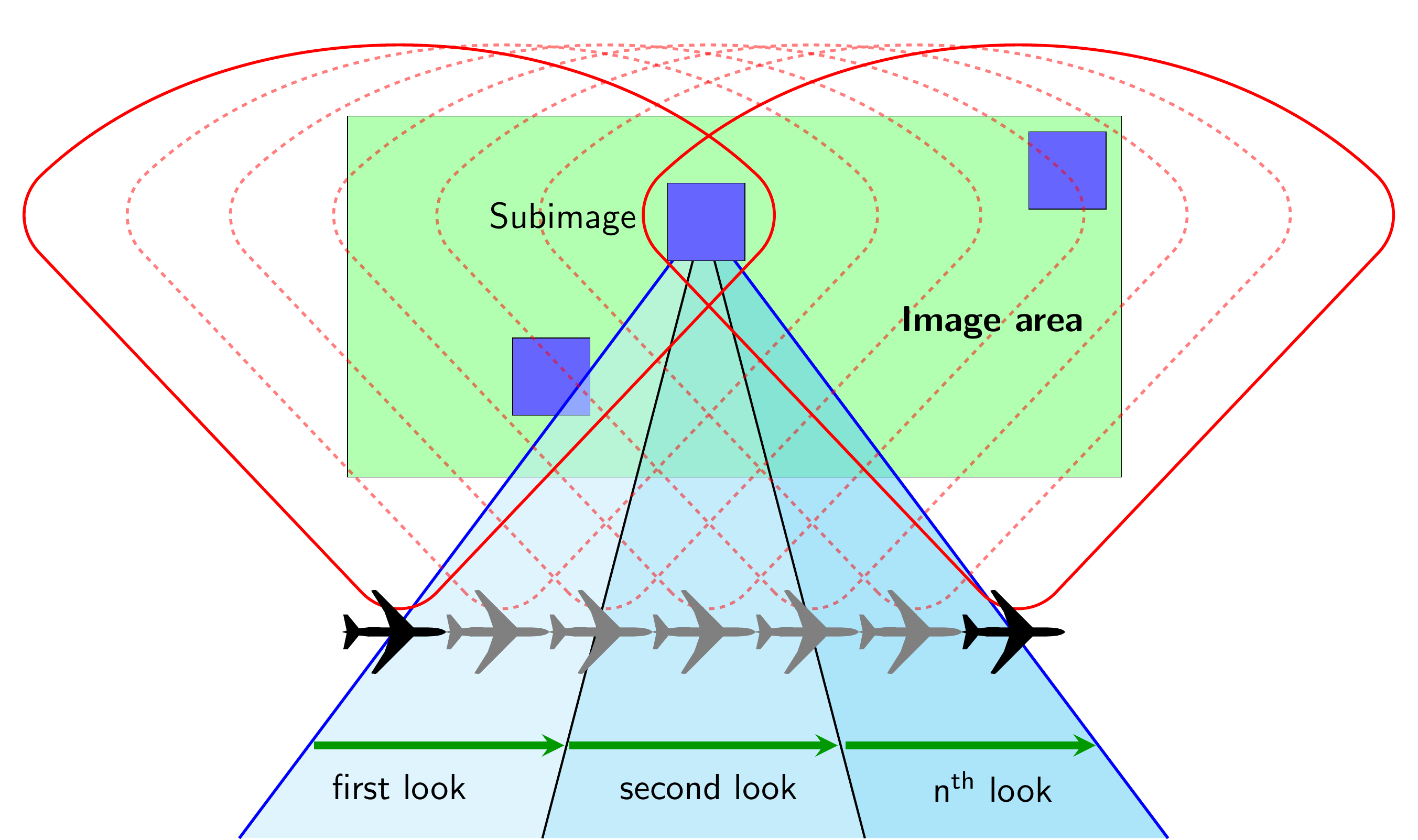}
\caption{\small Constant integration angle and multi-look SAR image formation 
via backprojection.}
\label{fig1}
\end{figure}

Figure~\ref{fig1} shows a typical side-looking SAR geometry
where a vehicle or an airborne-based radar transmits wideband signals to
the imaging area and measures the backscatter signals as it moves along an
aperture. Let $s_k$ be the received range-compressed data received at
$k^{\text{th}}$ aperture; the backprojection algorithm~\cite{nguyen1998ultra}
computes the value for each SAR image pixel from the imaging area as follows:
\begin{equation}
    P_i = \sum_{k = N_{i_1}}^{N_{i_2}}w_k * s_k(f(i, k)),
\end{equation}
where $N_{i_2} - N_{i_1}+1$ is the total number of aperture records used for the
coherent integration, $w_k$ is the weighting value at the $k^{\text{th}}$
aperture position, and $f(i, k)$ is the shift index to the signal $s_k$ for
$i^{\text{th}}$ pixel at the $k^{\text{th}}$ aperture position. For a typical
SAR image with $0^{\circ}$ aspect angle, $N_{i_1}$ and $N_{i_2}$ correspond to
the angle values of $-\alpha/2$ and $\alpha/2$ to form a SAR image with an
integration angle of $\alpha$. Note that for a true constant integration angle
SAR image formation, $N_{i_1}$ and $N_{i_2}$ are computed for every pixel of the
SAR image. However, for computational efficiency, a large image area is divided
into smaller subimages. For each subimage, SAR image formation is computed using
the $N_{i_1}$ and $N_{i_2}$ values derived from the geometry of the center of
the subimage and the radar aperture. To exploit the aspect dependence
information of target, instead of forming a single SAR image with $0^{\circ}$
aspect angle as described above with a single sector $[-\alpha/2, \alpha/2]$, we
form multiple sectors to generate the corresponding SAR images at different
aspect angles. For example, in our \textit{consecutive multi-look experiment},
three SAR images are formed: left side image that covers sector $[-\alpha/2,
\alpha/6]$, broadside image that covers sector $[-\alpha/6,
\alpha/6]$, and right side image at sector {$[\alpha/6, \alpha/2]$}.

To achieve a constant cross-range resolution SAR image, a large image area is
also divided into smaller subimages. Each subimage is formed using a different subset
of aperture. For a constant integration angle, subimages at farther range would
be integrated using a longer aperture than near-range subimages. The subimages
are then mosaicked together to form a single large image that covers the area of
interest. In consecutive multi-look SAR image processing mode, instead of
generating a single SAR image, multiple SAR images are formed at different
viewing angles to exploit the aspect angle dependent information from
targets. Thus, each aperture for each subimage is further divided into smaller
segments (either disjoint or overlapped) that are used to form consecutive
multi-look SAR images as illustrated in Figure~\ref{fig1}. The
aspect angle dependence features from targets have been exploited in {past}
research before using different
techniques~\cite{damarla2001detection,runkle2001multi,nguyen1998ultra}.

\subsection{Closely related works and motivation}
\label{ssec:related}

{UWB radar techniques have recently attracted increasing attention in the
area of penetration and object detection, thanks to their usage in
security applications and surveillance systems~\cite{anabuki2017ultrawideband}.
T. Sakamoto \etal~\cite{sakamoto2016fast} proposed fast methods {for}
   ultra-wideband (UWB) radar imaging that can be applied to a moving target.
   The technology has been also applied to 3-D imaging applications~\cite{yamaryo2018range}, human
   posture~\cite{kiasari2014classification}, human
   activity~\cite{bryan2012application}, vital sign~\cite{liang2018through},
   and liquid material~\cite{wang2015classification} classification problems. In
   these papers, due to high dimensionality and small signal-to-ratio (SNR), the
   signals need to be preprocessed, e.g., dimensionality reduction and
   background subtraction, before being used to train a classifier. It has been
   shown that support vector machines usually provide the best
   performance~\cite{bryan2012application,wang2015classification}. It is worth
   noting that in the aforementioned applications, objects are usually big
   (human) and captured from a relatively small distance. On the contrary,
   objects in our problem are relatively small and quite far from the radar. }

In this paper, we consider the problem of discriminating and classifying buried
targets of interest (metal and plastic mines, 155-mm unexploded ordinance [UXO],
etc.) from other natural and manmade clutter objects (a soda can, rocks, etc.)
in the presence of noisy responses from the rough ground surfaces for
low-frequency UWB 2-D SAR images. For classification problems, sparse
representation-based classification~\cite{Wright2009SRC} (SRC) has been
successfully demonstrated in other imagery domains such as medical image
classification~\cite{vu2015dfdl, Srinivas2013, vu2016tmi, Srinivas2014SHIRC},
hyperspectral image classification~\cite{sun2015task, sun2014structured,
chen2013hyperspectral}, high-resolution X-band SAR image
classification~\cite{zhang2012multi}, video anomaly detection
\cite{mo2014adaptive}, and several others
\cite{vu2016icip,Mousavi2014ICIP, vu2016fast, srinivas2015structured,
zhang2012joint, dao2014structured, dao2016collaborative, van2013design}.
However, in the low-frequency RF UWB SAR domain, although we have studied the
feasibility of using SRC for higher-resolution 3-D down-looking SAR
imagery~\cite{lamnguyen2013}, the application of SRC to low-frequency UWB 2-D
SAR imagery has not been studied to date due to the aforementioned
low-resolution issue. In this paper, we generalize the traditional SRC to
{address} target classification using either a single channel (radar
polarization) or multiple channels of SAR imagery. Additionally, we further
propose a novel discriminative tensor sparsity framework for multi-look
multi-channel classification problem, which is naturally suitable for our
problem. In sparse representations, many signals can be expressed by a linear
combination of a few basic elements taken from a ``dictionary''. Based on this theory,
SRC~\cite{Wright2009SRC} was originally developed for robust face recognition.
The main idea in SRC is to represent a test sample as a linear combination of
samples from the available training set. Sparsity manifests because most of the
nonzero components correspond to basic elements with the same class as the test sample.

Multi-channel SRC has
been {investigated} before in medical images~\cite{Srinivas2013, Srinivas2014SHIRC}.
In these papers, one dictionary for each channel is formed from training data
with locations of all channels of one training point being the same in all
dictionaries. Then intuitively, when sparsely encoding each channel of a new
test point using these dictionaries, we obtain sparse codes whose active
(nonzero) elements tend to happen at the same locations in all channels. In
other words, active elements are simultaneously located at the same location
across all channels. This intuition is formulated based on $l_0$ pseudo-norm,
which is solved using a modified version of simultaneous orthogonal matching
pursuit (SOMP)~\cite{tropp2006algorithms}. The cost function is nonconvex, and
hence, it is difficult to find the global solution. Furthermore, when more
constraints involved, there is no straightforward way to extend the algorithm.
In this paper, we proposed another way of formulating the simultaneity
constraint based on the $l_{12}$ norm, which enforces the row sparsity of the
code matrix (in tensor form, we call it \textit{tube sparsity}). The newly
convex optimization problem can be solved effectively using the fast iterative
shrinkage thresholding algorithm (FISTA)~\cite{beck2009fast}. We also propose
other different tensor sparsity models for our multi-channel classification
problems.

It has been shown that learning a dictionary from the training samples instead
of concatenating all of them as a dictionary can further enhance performance of
sparsity-based methods. On one hand, the training set can be compacted into a smaller dictionary, reducing computational burden at the test time. On the other
hand, by using dictionary learning, discriminative information of different
classes can be trained via structured discriminative constraints on the
dictionary as well as the sparse tensor code. A comprehensive study of
discriminative dictionary learning methods with implementations is presented
at~\cite{vu2016fast,vu2017dictol}. These dictionary learning methods, however, are all
applied to single-channel problem where samples are often represented in form of
vectors. While multi-channel signals can be converted to a long vector by
concatenating all channels, this trivial modification not only leads to the
\textit{curse of dimensionality} of high-dimensional space, but also possibly
neglects cross-channel information, which might be crucial for classification. In
this paper, we also propose a method named TensorDL, which is a natural
extension of single-channel dictionary learning frameworks to multi-channel
dictionary learning ones. Particularly, the cross-channel information will be
captured using the aforementioned simultaneity constraint.

Naturally, when a radar carried by a vehicle or aircraft moves around an
object of interest, it can capture multiple consecutive views of that object
(see Fig.~\ref{fig1}). Consequently, if the multi-look
information is exploited, the classification accuracy will be improved. While
the multi-look classification problem has been
approached before by SRC-related methods~\cite{zhang2012multi, Mousavi2014ICIP}, none of these works
uses the relative continuity of different views. We propose a framework to
intuitively exploit this important information. More importantly, the
optimization problem corresponding to this structure can be converted to the
simultaneous sparsity model by using an elegant trick {that we call} ShiftSRC.
Essentially, a tensor dictionary is built by circularly shifting an appropriate
amount of a single-channel dictionary. When we sparsely code the multi-look
signals using this tensor dictionary, the tensor sparse code becomes tube
sparsity.

\subsection{Contributions}
{The main contributions of this paper are as follows:}

\begin{enumerate}

    \item \textbf{A framework for simultaneously denoising and classifying 2-D
    UWB SAR imagery}\footnote{The preliminary version of this work was presented
    in IEEE Radar Conference, 2017~\cite{vu2017tensor}}. Subtle features from
    targets of interest are directly learned from their SAR imagery. The
    classification also exploits polarization diversity and consecutive aspect
    angle dependence information of targets.


    \item \textbf{A generalized tensor discriminative dictionary learning} 
    (TensorDL) is also proposed when more training data involved. These
    dictionary learning frameworks are shown to be robust even with high levels
    of noise. 

    \item \textbf{A relative SRC framework} (ShiftSRC) is proposed to deal with
    multi-look data. Low-frequency UWB SAR signals are often captured at
    different views of objects, depending on the movement of the radar carriers.
    These signals contain uniquely important information of consecutive views.
    With ShiftSRC, this information will be comprehensively exploited.
    Importantly, a solution to the ShiftSRC framework can be obtained by an
    elegant modification on the training dictionary, resulting in a tensor
    sparse coding problem, which is similar to a problem proposed in 
    contribution 1).

\end{enumerate}

The remainder of this paper is organized as follows. Section II presents
different tensor sparsity frameworks and  the discriminative tensor dictionary
learning scheme for multi-channel classification problems. The ShiftSRC for
multiple-relative-look and solutions to all proposed frameworks are also presented
in this section. Section III shows extensive experimental results on a
simulated dataset for several scenarios. An experiment with a realistic dataset
is also included. Section IV concludes the paper.

\section{Sparse representation-based classification} 
\label{sec:sparse_representation_based_classifications}

\subsection{Notation} 
\label{sub:notation}
Scalars are denoted by italic letters and may be either lower or uppercase,
e.g., $d, N, k$. Vectors and matrices are denoted by bold lowercase ($\mathbf{x,
y}$) and bold upper case ($\mathbf{X, Y}$), respectively. In this paper, we also
consider 3-D tensors (tensors for short) whose dimensions are named \textit{row,
column,} and \textit{channel}. A tensor with only one column will be denoted by
a bold, lowercase, calligraphic letter ($\bx, \by$). Tensors with more than one
column will be denoted by an bold, uppercase, calligraphic letters ($\bX, \bY,
\bD$).

For any tensor $\bM$, let $\bM^{(t)}$ be its $t$-th channel. For convenience,
given two tensors $\bM, \bN$, the tensor multiplication $\bP = \bM\bN$ is
considered channel-wise multiplication, i.e.,  $\bP^{(t)} = \bM^{(t)}\bN^{(t)}$.
For a tensor $\bM$, we also denote the sum of  square of all elements by
$\|\bM\|_F^2$ and the sum of absolute values of all elements by $\|\bM\|_1$.
Tensor addition/subtraction simply represents element-wise addition/subtraction.
Each target sample is represented by a UWB SAR image formed using either a
single (using co-pol) or multiple polarization (using both co-pol and cross-pol)
channels.  Thus, one target sample is denoted by  $\by \in \R^{d\times 1\times
T}$, where $d$ is the total number of image pixels and $T$ is the number of
polarization channels. A collection of $N$ samples is denoted by $\bY \in \R^{d\times N\times
T}$.

{Consider} a general classification problem with $C$ different
classes. Let $\bD_c (1 \leq c \leq C)$ be the collection of all training samples
from class $c$, $\bD_0$ be the collection of samples in the shared class, and
$\bD = [\bD_1, \dots, \bD_C, \bD_0]$ be the total dictionary with
the concatenation being done at the second dimension (column). In our problem,
the shared class can be seen as the collection of ground images.

\subsection{Classification scheme} 
\label{sub:classification_scheme}


\par

\par
{Using} the definition of tensor multiplication, a sparse representation of
$\by$ using $\bD$ can be obtained by solving
\begin{equation}
\label{eqn:general_opt}
    \bx = \arg\min_{\bx} \frac{1}{2} \|\by - \bD \bx\|_F^2 + \lambda\tF(\bx)
\end{equation}
\noindent where $\lambda$ is a positive regularization parameter and
$\tF(\bx)$ is a function that encourages {$\bx$ to be sparse}.
Denote by $\bx^i$ the sparse coefficient of $\by$ on $\bD_i$. Then, the
tensor $\bx$ can be divided into $C+1$ tensor parts
$\bx^1,\bx^2,\dots,\bx^C,\bx^0$.

After solving the sparse coding problem~\eqref{eqn:general_opt}, shared features
(grounds in our problem) are eliminated by taking $\bar{\by} = \by -
\bD_0\bx^0$. Then the identity of one sample $\by$ can be determined by the 
dictionary that provides the minimum residual:
\begin{equation}
    \text{identity}(\by) = \min_{i \in \{1, 2, \dots, C\}} \|\bar{\by} - \bD_i
    \bx^i\|^2_2 \end{equation}

\textit{Confuser detection:} In practical problems, the set of confusers is not
limited to the training set. A confuser can be anything that is not a target; it
can be solely the ground or a capture of an unseen object. In the former case,
the test signal can be well represented by using only the ground $\bD_0$, while in
the latter case, the sparse representation assumption is no longer valid.
Therefore, one test signal $\by$ is classified as a confuser if one {of}
following three conditions {is satisfied}: i) it is not sparsely interpreted by the
total dictionary $\bD$; ii) it has the most active elements in the sparse code
locating at $\bx^0$; and iii) it is similar to known confusers.

\subsection{Generalized sparse representation-based classification} 
\label{sub:general_sparse_representation_based_classifications}
\par

\begin{figure}[t]
    \centering
    \includegraphics[width = 0.5\textwidth]{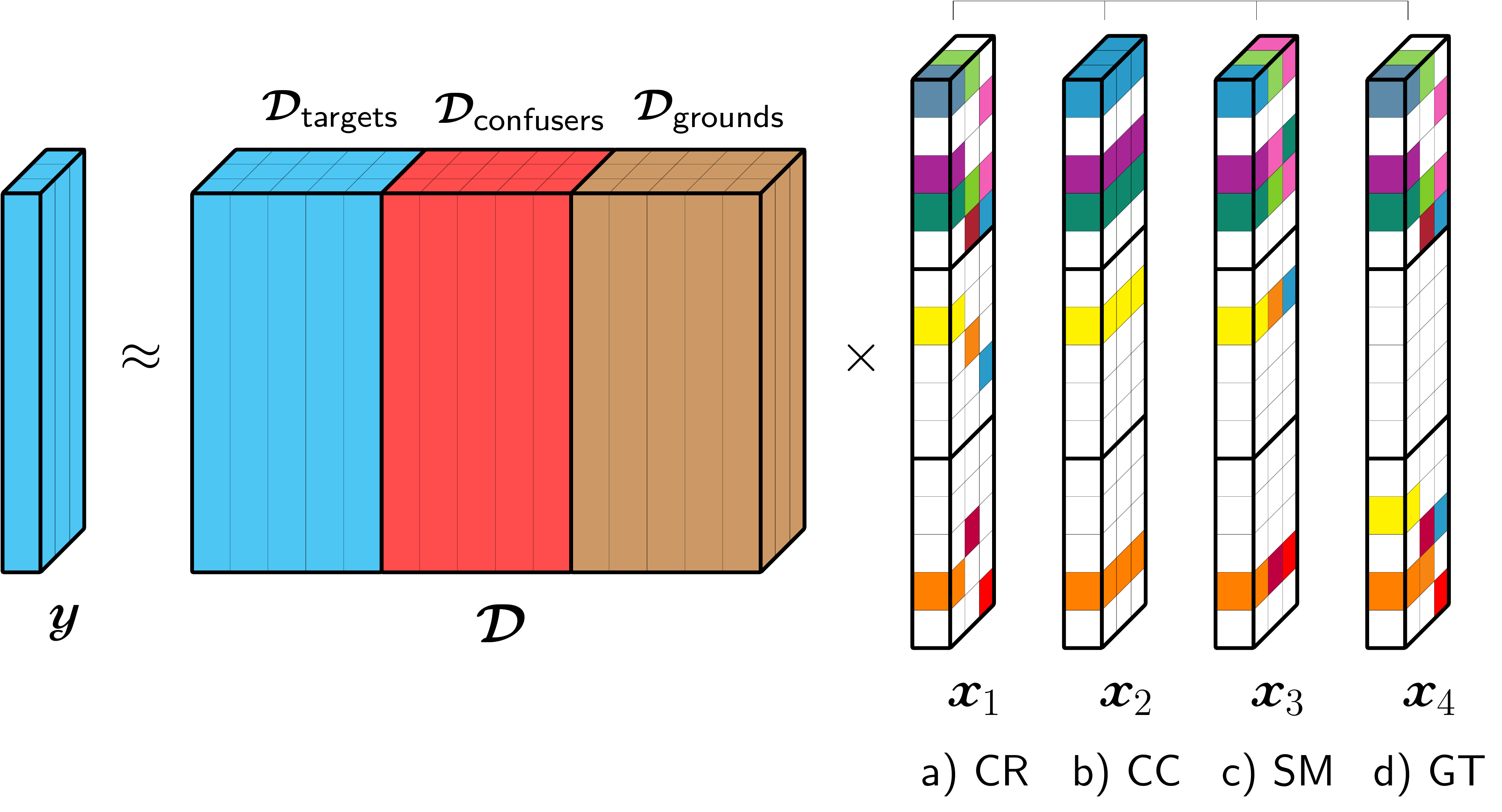} 
    \caption{\small  Different sparsity constraints on the coefficient tensor
    $\bx$.}
    \label{fig2}
\end{figure}



In SRC~\cite{Wright2009SRC} where only one channel is considered, $g(\bx)$ is
simply a function that forces sparsity as $l_0$- or $l_1$-minimization.
$l_1$-minimization is often used since it leads to a convex optimization problem
with tractable solutions. In the following, we present two natural extensions of
SRC for multi-channel cases. We also proposed two tensor sparse representation
methods that can enhance classification accuracy by exploiting
cross-channel information. These four generalized SRC methods are as follows.

\textit{a) Apply a sparsity constraint on each column of the coefficient
tensor $\bx$, no cross-channel constraint}.

We can see that this is similar to solving $T$ separate sparse coding
problems, each for a polarization channel. The classification rule is
executed based on the sum of all the squares of the residuals. We refer
to this framework as SRC-cumulative residual or SRC-CR (CR in the short
version). (See Figure~\ref{fig2}a with sparse tensor $\bx_1$). The
sparse code corresponding to the $t^{\text{th}}$ channel,
$\bxt$, is computed via the traditional $l_1$-minimization:
\begin{equation}
    \bxt = \arg\min_{\bxt} \frac{1}{2} \|\byt - \bDt\bxt\|_2^2 + \lambda\|\bxt\|_1
\end{equation}
which can be solved effectively using FISTA~\cite{beck2009fast},
ADMM~\cite{boyd2011distributed}, etc., algorithms or the SPAMS
toolbox~\cite{SPAMS}.

\textit{b) Concatenate all channels}.

The most convenient way to convert a multi-channel problem to a single-channel problem is
to concatenate all $T$ channels of one signal to obtain a long vector. By doing
so, we have the original SRC framework with $l_1$-norm minimization. After
solving the sparse coding problem, if we \textit{break} the long vectors and
rearrange them back to tensor forms, then the tensor sparse coefficients $\bx$
can be formed by replicating the one-channel sparse coefficient at all channel
(see Figure~\ref{fig2}b with sparse tensor $\bx_2$). From this tensor
viewpoint, the code tensor $\bx$ will have few active ``tubes''; moreover, all
elements in a tube are the same. We refer to this framework as SRC-concatenation
or SRC-CC (CC in the short version).

The optimization problem in SRC-CC and its solution are very straightforward.
First, we stack all channel dictionaries into a long one: $\hat{\bD} =
[\bD^{(1)}; \bD^{(2)}; \dots; \bD^{(T)}]$ (symbol `;' represents the
concatenation in the first dimension). Then for every test signal, we also stack
all of its channels to form a long vector: $\hat{\by} = [\by^{(1)}; \by^{(2)};
\dots; \by^{(T)}]$. The optimization problem \eqref{eqn:general_opt} becomes the
traditional $l_1$ regularization problem:
\begin{equation}
    \bx = \arg\min_{\bx}\frac{1}{2} \|\hat{\by} - \hat{\bD}\bx\|_2^2 + \lambda\|\bx\|_1
\end{equation}
then also can be solved by aforementioned methods.

\textit{c) Use a simultaneous model}.

\par Similar to the SHIRC model proposed in~\cite{Srinivas2013,
Srinivas2014SHIRC}, we can impose one constraint on active elements of tensor
$\bx$ as follows: $\bx$ also has few nonzero tubes as in  SRC-CC; however,
elements in one active tube are not necessarily the same. In  other words, the
locations of nonzero coefficients of training samples in the  linear combination
exhibit a one-to-one correspondence across channels. If the  $j$-th training
sample in $\bD^{(1)}$ has a nonzero contribution to $\by^{(1)}$,  then for $t
\in \{2, \dots, T\}$, $\byt$ also has a nonzero contribution from the $j$-th
training sample in $\bDt$. We refer to this framework as SRC-Simultaneous  or
SRC-SM (SM in the short version). (See Figure~\ref{fig2}c with  sparse
tensor $\bx_3$). To achieve this requirement, we can impose on the tensor
$\bx_3$ (with one column and T channels) the $l_{1,2}$-minimization constraint,
which is similar to the row-sparsity constraint applied on matrices
in~\cite{zhang2012multi}.

\par \textbf{Remarks:} While SHIRC uses $l_0$-minimization on $\bx$ and
applies the modified SOMP~\cite{tropp2006algorithms}, our proposed SRC-SM
exploits the flexibility of $l_{1,2}$-regularizer since it is convex, and easily
modified when more constraints are present (e.g., non-negativity). In
addition, it is more efficient especially when dealing with problems of
multiple samples at input.

Concretely, the optimization problem of SRC-SM can be written in the form
\begin{equation}
    \label{eqn:src_sm}
    \bx = \arg\min_{\bx} \frac{1}{2} \|\by - \bD\bx\|_F^2 + 
            \lambda \sum_{k=1}^K \|\vec(\bx_{k::})\|_2
\end{equation}
where $\bx_{k::}$ denotes the $k^{\text{th}}$ \text{tube} of the tensor code
$\bx$ and $K$ is the total column of the dictionary $\bD$, and $\vec(\bx_{k::})$ 
is the vectorization of the tube $\bx_{k::}$. This problem is similar to the 
joint sparse representation (JSRC) problem proposed in~\cite{zhang2012multi}
except that SRC-SM enforces tube-sparsity instead of the row-sparsity.
Details of the algorithm {that solves~\label{eqn:src_sm} are} described in Section~\ref{sec:solution}.

\textit{d) Use a group tensor model.}

Intuitively, since one object is ideally {represented} by a linear combination of
the corresponding dictionary and the shared dictionary, it is highly {likely}
that number of active (tensor) parts in $\bx$ is small, i.e., most of $\bx^1,
\dots, \bx^C, \bx^0$ are zero tensors. This suggests us a \textit{group tensor}
sparsity framework as an extension of~\cite{Yu2011} that can improve the
classification performance, which is referred to as SRC-GT (GT in the short
version). The visualization of this idea is shown in
Figure~\ref{fig2}d.

The optimization problem of SRC-GT is similar to \eqref{eqn:src_sm} with a slight
difference in the grouping coefficients:
\begin{equation}
    \label{eqn:src_gt}
    \bx = \arg\min_{\bx} \frac{1}{2} \|\by - \bD\bx\|_F^2 + 
            \lambda \sum_{c=1}^{C+1} \|\vec(\bx^c)\|_2
\end{equation}
where $C+1$ is the total number of classes (including the shared ground class), 
and $\vec(\bx^c)$ is the vectorization of the group tensor $\bx^c$. Solution to 
this problem will be discussed next.



The overall algorithm of generalized SRC applied to multi-channel signals in
the presence of a shared class is shown in Algorithm~\ref{alg:generalSRC}.

\begin{algorithm}[!t]
    \caption{Generalized SRC with a shared class}
    \label{alg:generalSRC}
    \begin{algorithmic}
    \Function {identity$(\by)$ = \\~~~~~~~~~~$~~~~~~~~~~\text{GENERALIZED\_SRC}(\by, \bD, \lambda, \tF(\bullet), \varepsilon, \tau)$ }{}
    \State \textbf{INPUT}: \\
    $\by \in \R^{d \times 1 \times T}$ -- a test sample; \\
    $\bD = [\bD_1, \bD_2, \dots, \bD_C, \bD_0] \in \R^{d\times K \times T}$ -- the total dictionary with the shared dictionary $\bD_0$; \\
    $\tF(\bullet)$ -- the sparsity constraint imposed on sparse codes.\\
    $\lambda \in \R^{+}$ -- a positive regularization parameter;\\
    $\varepsilon, \tau$ -- positive thresholds.
    \State \textbf{OUTPUT}: the identity of $\by$.
    \State 1. Sparsely code $\by$ on $\bD$ via solving:
   \begin{equation}
       {\bx} = \arg\min_{\bx} \{\|\by - \bD\bx\|_F^2 + \lambda\tF(\bx)\}
       \label{eqn: src}
   \end{equation}
   \State 2. Remove the contribution of the shared dictionary: $$\displaystyle
   \bar{\by} = \by - \bD_0\bx^0.$$
\State 3. Calculate the class-specific residuals :
{ $$\displaystyle r_c = \|\bar{\by} - \bD_c\bx^c\|_2, \forall c = 1, 2, \dots, C.$$}
    \State 4. Decision:
    \If{$\displaystyle\min_c(r_c) > \tau$ ({\it an unseen object}) \textbf{or} $\|\bar{\by}\|_2 < \varepsilon$ ({\it a ground})}
        \State $\by$ is a confuser.
    \Else
        \State $\displaystyle \text{identity}(\by) = \arg\min_{c}\{r_c\}$
   \EndIf
    \EndFunction
    \end{algorithmic}
\end{algorithm}


\subsection{Dictionary learning for tensor sparsity} 
\label{sub:dictionary_learning_for_tensor_sparsity}
As a natural extension, we can extend the tensor sparsity models to
dictionary learning ones. Most of dictionary learning methods focus on a
single-channel signal, which is not suitable for models with cross-channel
information. In this work, we extend single-channel dictionary learning
methods to multi-channel dictionary ones by applying aforementioned tensor
sparsity constraints.

\def\sbx{\mathbf{x}}
\def\sbX{\mathbf{X}}
\def\sbD{\mathbf{D}}
\def\sbY{\mathbf{Y}}

In single-channel, most of discrimination dictionary learning methods, such as 
FDDL~\cite{yang2014sparse}, DLSI~\cite{ramirez2010classification}, 
DFDL~\cite{vu2016tmi}, LRSDL~\cite{vu2016fast}, etc., have a cost function {that is of the form}
\begin{equation}
    \bar{J}_{\sbY}(\sbD, \sbX) = \bar{f}_{\sbY}(\sbD, \sbX) + \lambda \bar{g}
    (\sbX)
\end{equation}
where $\bar{g}(\sbX)$ is a function enforcing the sparsity of $\sbX$, and $
\bar{f}_{\sbY} (\sbD, \sbX)$, which includes fidelity and discriminant terms, is 
a function of $\sbD, \sbX$ and depends on the training samples $\sbY$.

One straightforward extension of these single-channel models to a multi-channel
case is to apply the first term $f_{\sbY}(\sbD, \sbX)$ to each channel and 
\textit{join} all channels by a sparsity constraint represented by $g(\sbX)$.
Naturally, $g(\sbX)$ can be one of four presented cross-channel sparsity
constraints. Concretely, the overall cost function would be in the form
\begin{equation}
    {J}_{\bY}(\bD, \bX) = {f}_{\bY}(\bD, \bX) + \lambda g(\bX)
\end{equation}
where $f_{\bY^{(t)}}(\bD^{(t)}, \bX^{(t)}) = \bar{f}_{\bY^{(t)}}(\bD^{(t)}, \bX^
{(t)})$ and $g(\bX)$ is one of \{CR, CC, SM, GT\} sparsity constraints. 

In this paper, we particularly focus on extending FDDL~\cite{yang2014sparse} to
the multi-channel case. FDDL is a special case of LRSDL~\cite{vu2016fast} without
an explicit shared dictionary. FDDL is chosen rather than
LRSDL since in our problem, the shared dictionary, e.g., grounds, is already
separated out. We also adopt  fast and efficient algorithms proposed in the
dictionary learning toolbox DICTOL~\cite{vu2017dictol} to update each channel of
the dictionary $\bD$. Also, the sparse code tensor $\bX$ is updated using
FISTA~\cite{beck2009fast} algorithm, which is discussed in Section
\ref{sec:solution}.

The proposed cross-channel dictionary learning method is named TensorDL suffixed
by CR, CC, SM, or GT when different sparsity constraints are applied on $\bX$.

\begin{figure}[t]
    \centering
    \includegraphics[width = 0.48\textwidth]{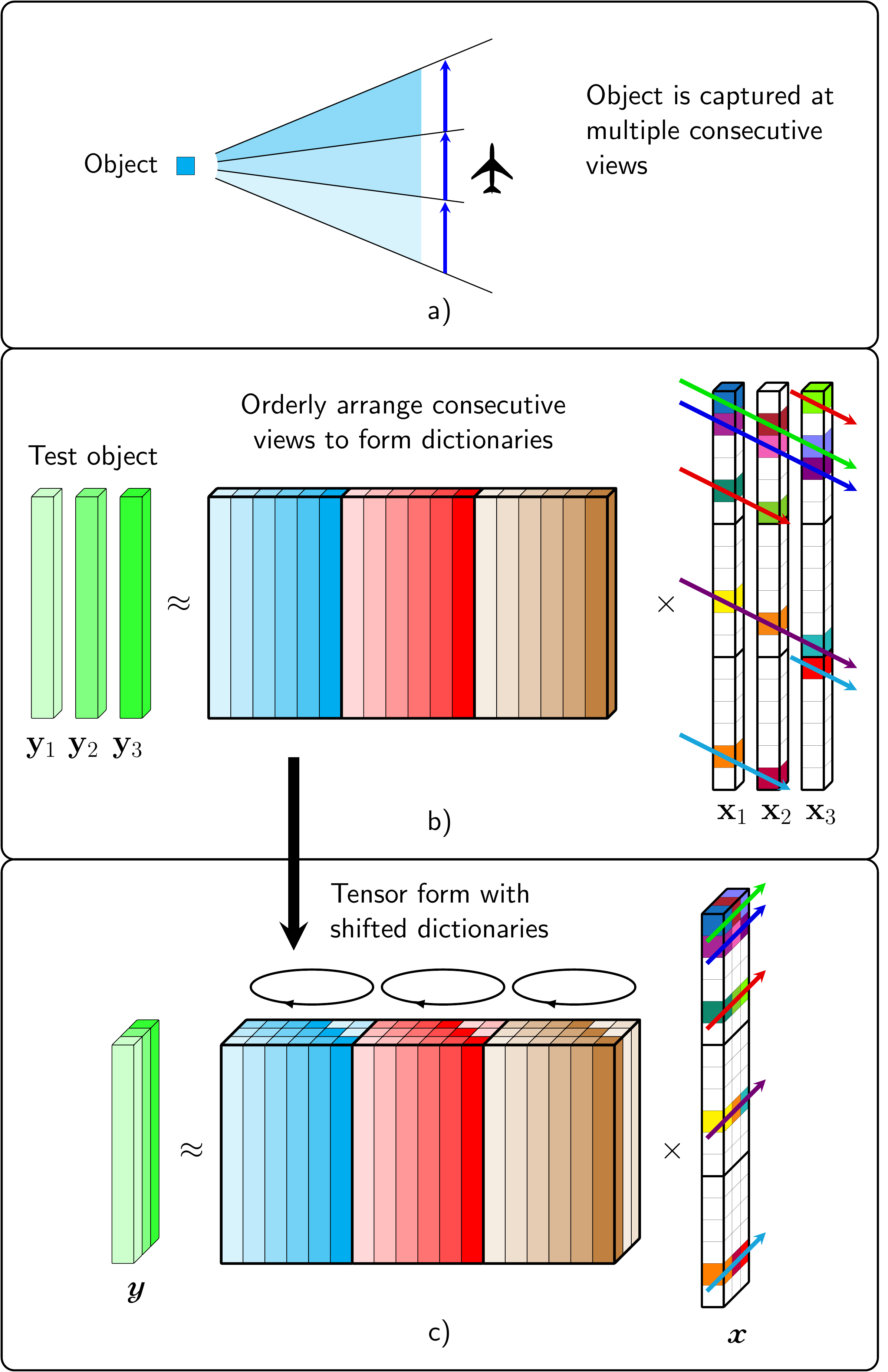}
    \caption{\small Tensor sparsity with relative views.}
    \label{fig3}
\end{figure}

\subsection{Tensor sparsity with multiple relative looks} 
\label{sub:tensor_sparsity_with_multi_relative_views}

\def\sby{\mathbf{y}}

In realistic situations, objects might be captured at different angles instead
of only one. Moreover, these are often consecutive angles that a plane or a
vehicle can capture (see Figure~\ref{fig3}a left) while moving
around objects. Based on this observation, in the training phase, we collect
data from different views of objects, as in Figure~\ref{fig3}a
right, and orderly arrange them into dictionaries for each object, as
illustrated in Figure~\ref{fig3}b. The test objects, which are
often captured at different relative views $\sby_1, \sby_2, \sby_3$, are then
sparsely coded by the whole dictionary.

Intuitively, if $\sbx_1$ is the sparse code of the first view $\sby_1$ with only
few active elements, then the sparse code $\sbx_2$ of the next view $\sby_2$
will be active at locations shifted by one. Similarly, active locations in
$\sbx_3$ of the third view will be shifted by two compared to $\sbx_1$, and so
on. In this case, active coefficients form a ``stair'', as illustrated in
Figure~\ref{fig3}b right.

The sparse coding problem with the ``stair'' sparsity is a novel problem and has
{not} been addressed before. In this paper, we propose a method called ShiftSRC
to convert this problem to a {previously known} problem. The central idea is that if we
stack all views into a tensor and ``circularly shift'' the ordered dictionary by
one to the left at each view, then the obtained tensor code will be a tensor
with active elements forming a ``tube'' sparsity (see
Figure~\ref{fig3}c). The solution to this problem is similar to the
solution of SRC-SM as stated in this paper.


\subsection{{Solution to optimization problems}} 
\label{sec:solution}

{ Both optimization problems \eqref{eqn:src_sm} and
\eqref{eqn:src_gt} have the form
\begin{equation}
    \label{eqn:common_fista}
    \bx = \arg\min_{\bx}\{F(\mathbcal{x}) \equiv f(\bx) + \lambda g(\bx)\},
\end{equation}
where
\begin{itemize}
    \item $g(\bx)$ is sum of norm 2, then it is a continuous convex function
    and {\em nonsmooth}.
    \item $f(\bx) = \frac{1}{2} \|\by - \bD\bx\|_F^2$ is a continuous convex
    function of the type $C^{1, 1}$, i.e., continuously differentiable with a
    Lipschitz continuous gradient $L$:
    \begin{equation*}
        \|\nabla f(\mathbcal{x}_1) -\nabla f(\mathbcal{x}_2)\|_F \leq L
        \|\bx_1 - \bx_2\|_F ~~ \text{for every}~ \bx_1, \bx_2.
    \end{equation*}
\end{itemize}
}
We observe that, with these properties, the optimization problem
\eqref{eqn:common_fista} can be solved by FISTA~\cite{beck2009fast}. There
are three main tasks in FISTA:\\

\noindent 1. Calculating $\nabla f(\bx)$, which can be easily computed
as $$\nabla f(\bx) = \bD^T(\bD\bx - \by).$$\\
\noindent where each channel of $\bD^T$ is the transpose of the corresponding
channel of
$\bD$.

\begin{algorithm}[t]
\label{alg:LRSDLX}
    \caption{{Tensor sparse coding by FISTA\cite{beck2009fast}}}
    \begin{spacing}{1.3}
    \begin{algorithmic}
    \Function {${\bx}$ = TENSOR\_SC}{$\by, \bD, \bx_{\text{init}}, \lambda$}.
    \State 1. Calculate
    \begin{align*}
        \bA &= \bD^T\bD, ~~~\bb = \bD^T\by \\
        L & = \max_{t = 1, 2, \dots, T} (\lambda_{\max}(\bA^{(t)}))
    \end{align*}
    \State 2. Initialize $\bx_0 = \bx_{\text{init}}$, $\bz_1 = \bx_0$, $j = 1, t_1 = 1$
    \While {non convergence and $j < j_{\max}$}
        \State 3. Calculate gradient: $\bg = \bA\bz_j - \by.$
        \State 4. Calculate $\bu = \bz_j - \bg/L.$
        \State 5. If SRC-SM, $\bx_j$ is the solution of \eqref{eqn:sm_norm2};
            if SRC-GT, $\bx_j$ is the solution of \eqref{eqn:gt_norm2}.
        \State 6. $t_{j+1} = (1 + \sqrt{1 + 4t_j^2})/2$

        \State 7. $\bz_{j+1} = \bx_j + \frac{t_j - 1}{t_{j+1}} (\bx_j - \bx_{j-1})$
        \State 8. $j = j + 1$

    \EndWhile
    \State 9. OUTPUT: $\bx = \bx_j$
    \EndFunction
    \end{algorithmic}
    \end{spacing}
\end{algorithm}

\noindent 2. Calculating a Lipschitz constant of $\nabla f(\bx)$. For our
function $f$, we can choose
\begin{equation}
    L = \max_{t = 1, 2, \dots, T} \left\{\lambda_{\max}\left((\bD^{(t)})^T\bDt\right)\right\}
\end{equation}
where $\lambda_{\max}$ is the maximum eigenvalue of a square matrix.

\noindent 3. Solving a suboptimization problem of the form
\begin{equation}
\label{eqn:fista_approx}
    \bx = \arg\min_{\bx} \left\{\frac{1}{2} \|\bx - \bu\|_F^2 + \eta g(\bx)\right\}
\end{equation}
with $\eta = \frac{\lambda}{L}$.

For SRC-SM, problem \eqref{eqn:fista_approx} has the form
\begin{eqnarray}
\nonumber
    \bx = \arg\min_{\bx} \left\{\frac{1}{2} \|\bx - \bu\|_F^2 + \eta \sum_{k=1}^K \|\vec(\bx_{k::})\|_2\right\} \\
    \label{eqn:sm_norm2} =  \arg\min_{\bx} \left\{ \sum_{k = 1}^K
    \left(\frac{1}{2} \|\bx_{k::} - \bu_{k::}\|_2^2 + \eta
    \|\vec(\bx_{k::})\|_2\right)\right\}
\end{eqnarray}
Each problem in \eqref{eqn:sm_norm2} is a minimum $l_2$-norm minimization with solution being
\begin{equation}
\label{eqn:sol_sm}
    \bx_{k::} = \max\left\{1 - \frac{\eta}{\|\vec(\bu_{k::})\|_2}, 0\right\}\bu_{k::}, \forall k = 1, 2, \dots, K.
\end{equation}
Similarly, for SRC-GT, problem \eqref{eqn:fista_approx} can be written as
\begin{eqnarray}
\nonumber
    \bx = \arg\min_{\bx} \left\{\frac{1}{2} \|\bx - \bu\|_F^2 + \eta \sum_{c=1}^{C+1}\|\vec(\bx^c\|_2\right\} \\
    \label{eqn:gt_norm2}
    =  \arg\min_{\bx} \left\{ \sum_{c = 1}^{C+1} \left(\frac{1}{2}
        \|\bx^c - \bu^c\|_2^2 + \eta \|\vec(\bx^c)\|_2\right)\right\}
\end{eqnarray}
with solution being:
\begin{equation}
\label{eqn:sol_gt}
    \bx^c = \max\left\{1 - \frac{\eta}{\|\vec(\bu^c)\|_2}, 0\right\}\bu^c,
    \forall c = 1, \dots, C+1.
\end{equation}

\begin{figure}
\centering
\includegraphics[width = .45\textwidth]{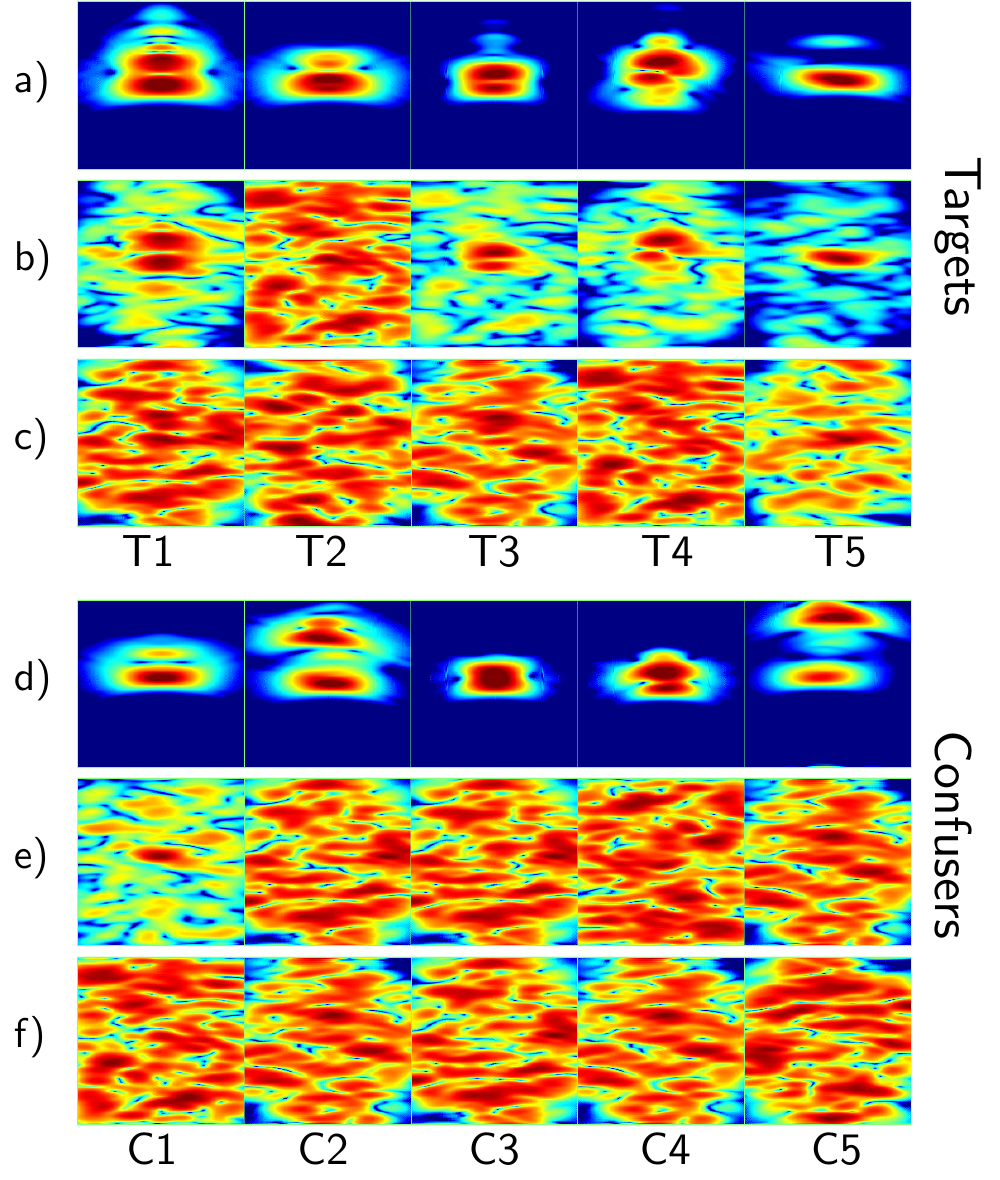}
\caption{Sample images of five targets and five clutter objects. T1 = M15
anti-tank mine, T2 = TM62P3 plastic mine, T5 = 155-mm artillery shell, C1 =
soda can, C2 = rocks, C3 = rocks, C4 = rocks, C5 = rocks. a) Targets under
smooth ground surface. b) Targets under rough ground surface (easy case,
scale = 1). c) Targets under rough ground surface (hard case, scale = 5). d)
Confusers under smooth ground surface. e) Confusers under rough ground
surface (easy case, scale=1). f) Confusers under rough ground surface (hard
case, scale=5).}
\label{fig4}
\end{figure}

{A step by step description of SRC-SM and SRC-GT algorithms are}
given in Algorithm 2.



\section{Experimental results} 
\label{sec:experimental_results}

\begin{figure}
\centering
\includegraphics[width = .49\textwidth]{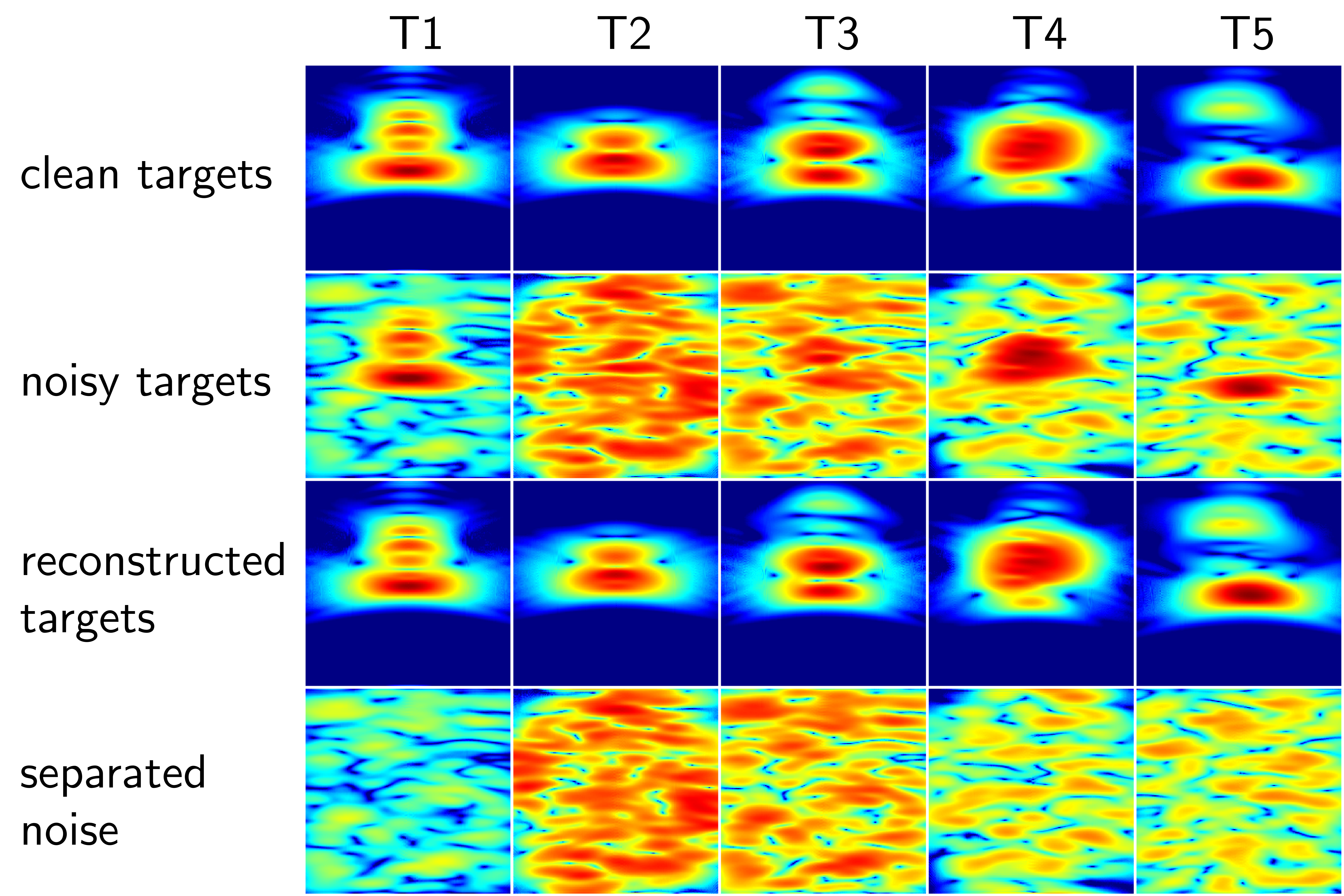}
\caption{Visualization of decomposed signals (HH polarization) after doing
sparse coding. Row 1: the original clean signals. Row 2: the corresponding
noisy signals with the interference of grounds. Row 3: reconstructed signal
after sparse coding. Row 4: separated noises.}
\label{fig5}
\end{figure}




In this section, we apply the proposed models to the problem of classifying
objects of interest. Extensive results are presented on simulated and real-life
datasets. A MATLAB toolbox for the tensor sparsity methods presented in this paper
is available at~\cite{vu2017tensorsparsity}.

\begin{table*}[]
\centering
\footnotesize
\caption{Overall average classification (\%) of different methods on different
polarization combinations, with or without the non-negativity constraint. }
\label{tab:tab1}
\begin{tabular}{c|c|c|c|c|c||c|c|c|c|c|c}
\cline{2-11}
                       & \multicolumn{5}{c||}{Separate targets} & 
                       \multicolumn{5}{c|}{All targets} &                                                                                                                    \\ \cline{1-11}
\multicolumn{1}{|l||}{} & VV   & HH   & VV+HH   & HH+HV  & ALL  & VV  & HH  & VV+HH  & HH+HV  & ALL  &                                                                                                                \\ \cline{1-11}
\multicolumn{1}{|l||}{SVM}    &62.17 &74.47 &65.00 &77.77 &70.33 &83.68 &89.36 &83.36  &66.36 &79.78 & \\ \hline \hline
\multicolumn{1}{|l||}{SRC-CR} &76.32 &83.43 &86.26 &85.93 &82.95 &86.22 &91.64
&87.14  &91.28 &90.64 &\multicolumn{1}{l|}{\multirow{4}{*}{\begin{tabular}[c]{@
{}l@{}}without \\ non-negativity \\ constraint\end{tabular}}} \\ \cline{1-11}
\multicolumn{1}{|l||}{SRC-CC} &76.32 &83.43 &79.47 &86.21 &79.93 &86.22 &91.64 &87.96  &92.98 &81.80 & \multicolumn{1}{l|}{}                                                                                              \\ \cline{1-11}
\multicolumn{1}{|l||}{SRC-SM} &76.32 &83.43 &81.07 &86.52 &85.87 &86.22 &91.64 &89.66  &95.12 &94.14 & \multicolumn{1}{l|}{}                                                                                              \\ \cline{1-11}
\multicolumn{1}{|l||}{SRC-GT} &69.79 &78.79 &77.83 &85.12 &81.75 &81.90 &88.40 &87.28  &91.86 &90.46 & \multicolumn{1}{l|}{}                                                                                              \\ \hline \hline
\multicolumn{1}{|l||}{SRC-CR} &{\bf 79.05} &{\bf89.41} &79.85 &88.52 &84.92 &
{\bf86.32} &{\bf92.98} &87.48  &94.86 &94.26 & \multicolumn{1}{l|}{\multirow{4}
{*}{\begin{tabular}[c]{@{}l@{}}with \\ non-negativity \\ constraint
\end{tabular}}}    \\ \cline{1-11}
\multicolumn{1}{|l||}{SRC-CC} &{\bf 79.05} &{\bf89.41} &83.99 &92.27 &85.11 &{\bf86.32} &{\bf92.98} &89.20  &93.82 &87.08 & \multicolumn{1}{l|}{}                                                                                              \\ \cline{1-11}
\multicolumn{1}{|l||}{SRC-SM} &{\bf 79.05} &{\bf89.41} &{\bf85.62} &{\bf90.85} &{\bf89.36} &{\bf86.32} &{\bf92.98} &{\bf90.00}  &{\bf96.82} &{\bf96.00} & \multicolumn{1}{l|}{}                                                                                              \\ \cline{1-11}
\multicolumn{1}{|l||}{SRC-GT} &75.55 &88.97 &81.85 &90.57 &87.28 &84.16 &90.32 &88.10  &94.58 &94.12 & \multicolumn{1}{l|}{}                                                                                              \\ \hline
\end{tabular}
\end{table*}
\subsection{Electromagnetic (EM) Simulation data}
SRC is applied to a SAR database consisting of targets (metal and plastic
mines, 155-mm unexploded ordinance [UXO], etc.) and clutter objects (a soda can,
rocks, etc.) buried under rough ground surfaces. The electromagnetic (EM) radar
data {is} simulated based on the full-wave computational EM method known as
the finite-difference, time-domain (FDTD) software~\cite{dogaru2010}, which was
developed by the U.S. Army Research Laboratory (ARL). The software was validated
for a wide variety of radar signature calculation
scenarios~\cite{dogaru2007,liao2012}. Our volumetric rough ground surface grid
-- with the embedded buried targets -- was generated by using the surface root-
mean-square (rms) height and the correlation length parameters. The  targets are
flush buried at 2-3 cm depth. In our experiments, the easiest case  of rough
ground surface in our experiments, the surface rms is 5.6 mm and the correlation
length is 15 cm. The SAR images of various targets and clutter objects are
generated from EM data by coherently integrated individual radar return signals
along over a range of aspect angles. The SAR images are formed using the
backprojection image formation~\cite{McCorkle1994} with an integration angle of
$30^{\circ}$. Figure~\ref{fig4}a shows the SAR images {(using vertical transmitter, vertical receiver -- VV --
polarization)} of some targets that are buried under a perfectly smooth ground
surface. Each target is imaged at a random viewing aspect angle and an
integration angle of $30^{\circ}$. Figures~\ref{fig4}b and
~\ref{fig4}c show the same targets as 
Figure~\ref{fig4}a, except that they are buried under a rough
ground surface (the easiest case corresponds to ground scale{/noise level}\footnote{{Note that a higher ground scale means more noisy images. Henceforth, in the text as well as figures we simply refer to ground scale as noise level for ease of exposition.}} = 1 and harder case
corresponds to noise level = 5). Similarly, Figures~\ref{fig4}d,
\ref{fig4}e, and~\ref{fig4}f show the SAR images of
some clutter objects buried under a smooth and rough surface, respectively. For
training, the target and clutter object are buried under a smooth surface to generate high signal-to-clutter ratio images. We include 12 SAR images
that correspond to 12 different aspect angles ($0^{\circ}, 30^{\circ},
\dots, 330^{\circ}$) for each target type. For testing, the SAR images of
targets and confusers are generated at random aspect angles and buried under 
rough ground surfaces. Various levels of ground surface roughness are simulated
by selecting different ground surface scaling factors when embedding the test
targets under the rough surfaces. Thus, the resulting test images are very noisy
with a very low signal-to-clutter ratio. 
{Each image is a polarization signal of object which is formed by one of transmitter-receiver setups: vertical-vertical (VV), horizontal-horizontal (HH), or horizontal-vertical (HV).}
Each data sample of one object is
represented by either i) one SAR image using data from one co-pol (VV, HH)
channel or ii) two or more images using data from co-pol (VV, HH) and cross-pol
(HV) channels. For each target type, we tested 100 image samples measured at
random aspect angles.

\subsection{Denoised signal visualization} 
\label{sub:denoised_signal_visualization}

We construct the dictionary $\bD = \bmt \bD_{t}, \bD_{c}, \bD_{c} \emt$ of 
all three polarizations, with $\bD_t, \bD_c, \bD_g$ being dictionaries of 
\textit{targets},
\textit{confusers}, and \textit{grounds}, respectively. The sub-dictionary $\bD_o
= \bmt \bD_t,
\bD_c \emt$ can be seen as the dictionary of the objects of interest. For
a noisy signal $\by$, we first solve the following problem:
\begin{equation}
    \bx = \arg\min_{\bx} \|\by - \bD\bx\|_F^2 + \lambda g(\bx)
\end{equation}
where $g(\bx)$ is the proposed SM constraint. The sparse tensor code $\bx$ is
then decomposed into two parts, $\bx^{o}$ and $\bx^{\text{g}}$. The latter can
be considered coefficients corresponding to the ground dictionary $\bD_g$. The
original signal can be approximately decomposed into two parts: $\bD_{g}\bx^{g}$
as separated noise, and $\bD_{o}\bx^o$ as the denoised signal. Visualization of
these signals are shown in Figure~\ref{fig5}. We can see that the
tensor framework successfully decompose{s} noisy signals into a clean part and a
ground signal. These results show the potential of the proposed tensor sparsity
frameworks for classification.

\subsection{Overall classification accuracy} 
\label{sub:overall_classification_accuracy_in_the_best_condition}
\begin{figure*}[t]
    \centering
    \includegraphics[width = 0.99\textwidth]{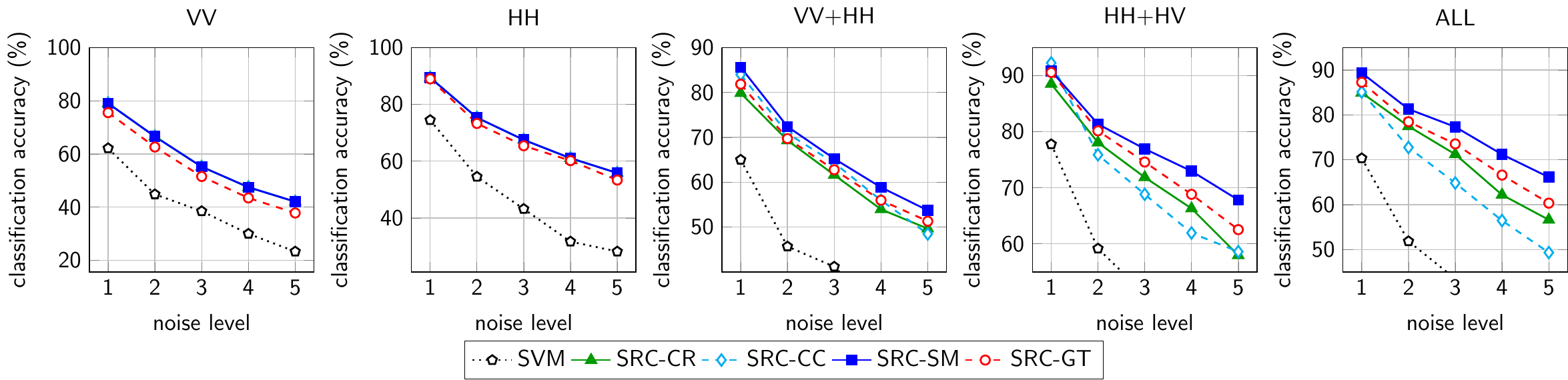}
    \caption{\small Classification accuracy (\%) with different {noise levels} and 
    different polarization combinations for the \textit{separate-target} scenario.}
    \label{fig6}
\end{figure*}

We apply four methods presented
in~\ref{sub:general_sparse_representation_based_classifications} to different
combinations of three polarizations: VV, HH, VV+HH, HH+HV, and VV+HH+HV (or
ALL), and also compare these results with those obtained by {a support vector machine} (SVM) using the libsvm
library \cite{CC01a}. {SVM was also applied to classify UWB
signals~\cite{wang2015classification,bryan2012application}}. The training set
comprises all five target sets, four out of five confuser sets (each time we
leave one confuser set out, which is meant to be unseen), and the ground set.
While all confuser sets can be seen as one class -- \textit{confuser class} --
there are two ways of organizing target sets. First, each of five targets is
considered one class in case the identity of each target is crucial (we name
this scenario \textit{separate-target} with five target classes and one confuser
class). Second, if we only need to know whether an object of interest is a
target or not, we can consider all five targets as one class and its
corresponding name is \textit{all-target} with one target class and one confuser
class. We also consider two families of the sparse coding problem, one with and
one without the non-negativity constraint on the sparse code $\bx$ in each of
the tensor sparse coding methods. Each experiment is conducted 10 times and
their results are reported as the average number. Parameters in each method are
chosen by a 5-fold cross-validation procedure. In this experiment, all test
samples are corrupted by small noise, i.e., the noise level is set to one. {In our experiments, we use overall classification accuracy as the performance metric, which computes percentage of correctly classified samples over total test samples across all classes.}

Overall classification accuracies of different methods on different
polarization combinations are reported in Table~\ref{tab:tab1}. From the table,
a few {inferences} can be made:

\begin{itemize}
    
    \item SRC-related methods with non-negative sparse coefficients perform
    better than those without this constraint\footnote{Based on this
    observation, from now on, all other results are implicitly reported with the
    non-negativity constraint.}. In addition, SVM is outperformed
    by all other methods in all tasks.

    \item SRC-SM provides the best results in all
    combinations for both the \textit{separate-target} and \textit{all-target}
    scenarios. The best accuracies in both cases are significantly high with
    slightly over 90\% in the six-class classification problem and nearly 97\%
    in the binary classification problem. 

    \item If only one polarization is used, SRC-CR, SRC-CC, and SRC-SM
    have identical results, since all of them are basically reduced to
    traditional SRC in this case. Additionally, these three methods slightly
    outperform SRC-GT in this scenario.

    \item If only one polarization can be obtained, HH always outperform{s} VV {and by a}
    significant {margin}. Additionally, the HH+VV combination often worsens
    the results {versus} using HH alone.

    \item If the same method is used on different combinations, the best results
    are mostly obtained by the combination of HH and HV polarizations in both
    the \textit{separate-target} and \textit{all-target} scenarios.

\end{itemize}

\subsection{{Effect of noise levels on overall accuracy}} 
\label{sub:effect_off_corruption_levels_on_overall_accuracy}

The results in the previous section are collected in the presence of small
corruption (noise level is only 1). In real problems, objects of interest are,
deliberately or not, buried under extremely rough surfaces in order to
\textit{fool} classifiers. In this section, we conduct an experiment to see how
each method performs when the level of corruption increases in the
\textit{separate-target} scenario.

Classification results of five different methods on different polarization
combinations and different noise levels are shown in
Figure~\ref{fig6}. First of all, similar trends to small corruption can
be observed in that SRC-SM shows best performance in all cases with bigger gaps
occurring at high levels of noise. In addition, of the four SRC-related methods,
SRC-CC is beaten by all three others when more than one polarization involved.
This can be explained by the fact that SRC-CC suffers from the \textit{curse of
dimensionality} when each sample is represented by concatenating long vectors.
It can also be seen that SRC-GT performs better than SRC-CR and SRC-CC in the
presence of multiple polarizations. Last but not least, the best classification
results can be seen at HH+HV and ALL among the five different combinations
considered.


\subsection{Effect of tensor dictionary learning on overall accuracy} 
\label{sub:effect_of_tensor_dictionary_learning_on_overall_accuracy}
We report the classification results for the
\textit{all-target} scenario with different noise levels. We also include the
results of experiments with discriminative tensor dictionary learning methods.
The results of HH+HV and ALL are reported, since they yield the best results, as
observed in previous sections.

\begin{figure}[t]
\centering
\includegraphics[width = .49\textwidth]{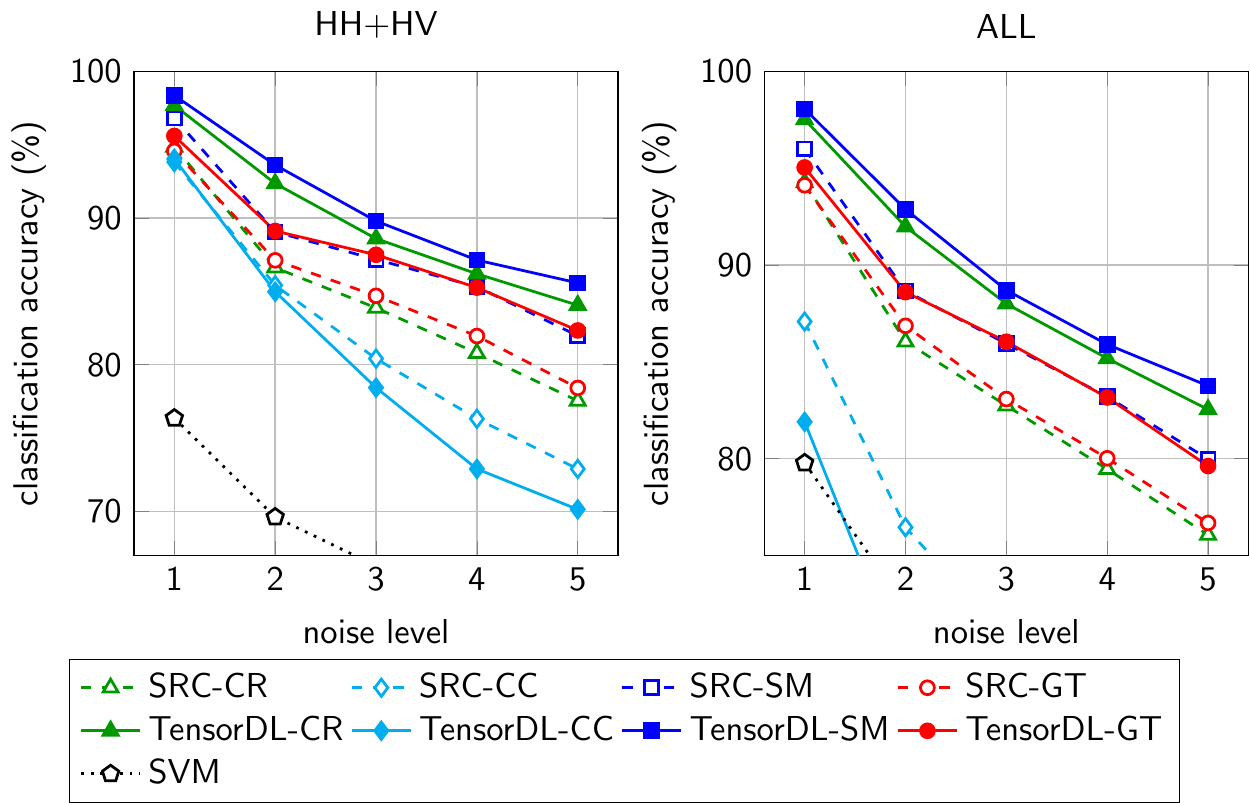}
\caption{\small Classification accuracy (\%) with different {noise levels},
    and a discriminative dictionary learning on the \textit{all-target}
    scenario.}
\label{fig7}
\end{figure}

The results of nine different methods are depicted in Figure~\ref{fig7}.
These methods include SVM (the dotted line), the four SRC-related methods (dashed
lines), and their corresponding tensor dictionary learning counterparts
(prefixed by {TensorDL}, solid lines). We can see that except for the
CC case, tensor dictionary learning methods outperform their
corresponding SRC with gaps widening as
the noise level increases. This observation confirms that tensor discriminative
dictionary learning methods indeed {provide} accuracy improvements. Methods with
the SM constraint {once again emerge} the winner in all noise levels, and the best accuracy numbers are observed using the HH+HV combination.


\subsection{Multiple-relative-look classification accuracy} 
\label{sub:multi_relative_look_classification_accuracy}


\begin{figure}[t]
\centering
\includegraphics[width = .45\textwidth]{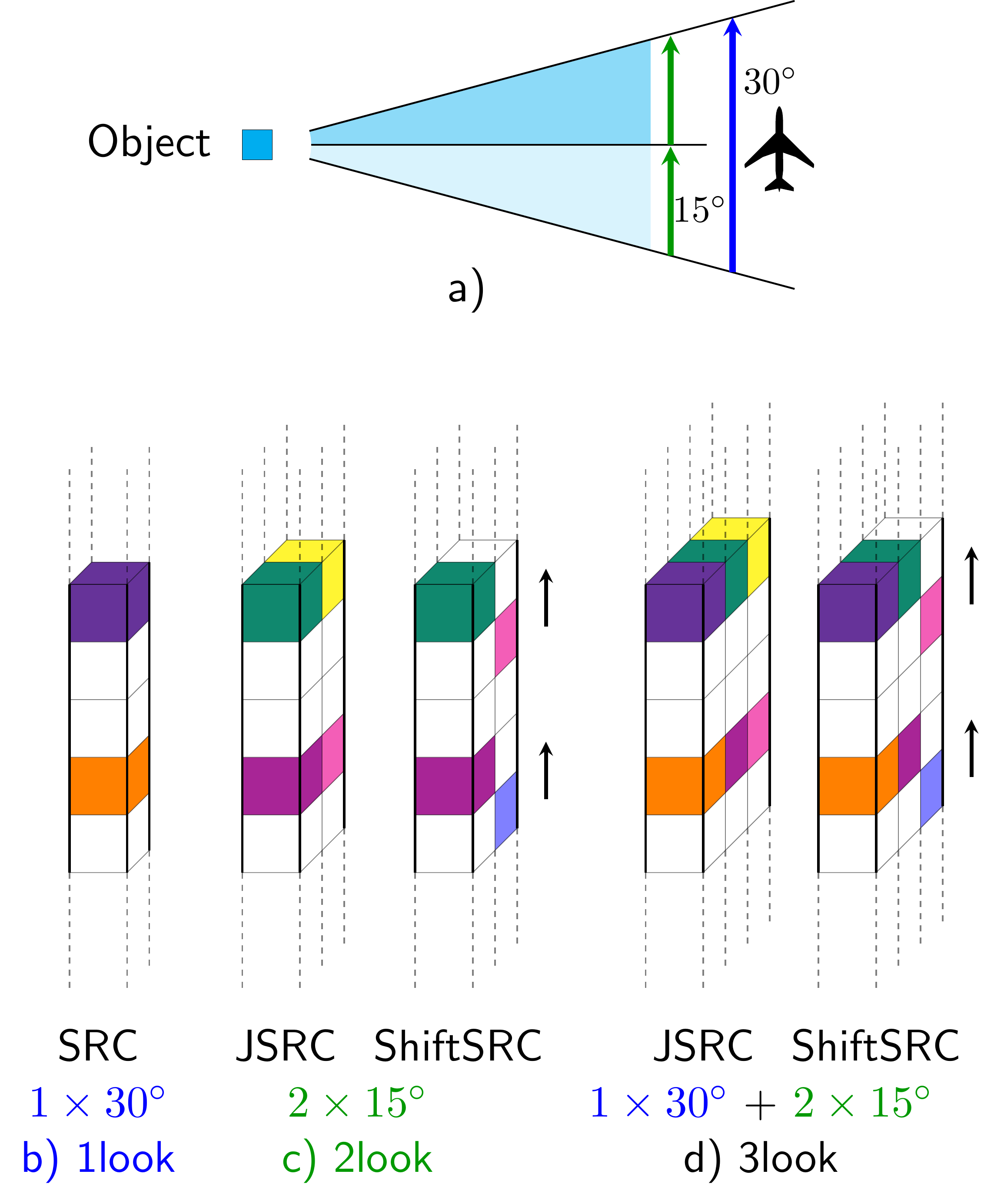}
\caption{\small Multi-relative-look experimental setup (a) and nonzeros locations
of sparse tensors in different scenarios (b, c, d).}
\label{fig8}
\end{figure}

We describe a key real-world scenario where multi-look signals of an object of
interest can be obtained. Figure~\ref{fig8}a depicts the relative
location of a radar carrier, a jet plane in this case, and an object of
interest. The plane moves around the object at an angle corresponding to the
blue arrow. One sample of object can be captured at this whole range, which can
be seen as one of its \textit{looks}. At the same time, the radar
can also capture multiple \textit{looks} of the object at smaller angles
represented by green arrows. By continuing considering even smaller angles, more
representatives of the object can be captured. These multiple views can provide
complementary information of the object, highly likely resulting in better
classification accuracy of the system.

For the training samples, for each object, we generate two sets of signals. Each
set contains samples captured after each $15^{\circ}$, and each set has total
of 24 views. Both sets start at the same relative angle but the first is
captured by an integration angle of $30^{\circ}$, the angle in the second set
is $15^{\circ}$. For the test samples, each object is represented by three
signals: one by an integration angle of $30^{\circ}$ and two others by
an integration angle of $15^{\circ}$, as depicted in Figure~\ref{fig8}a.
Similar to previous experiments, test samples are captured at random relative
angles. Ground samples are also simulated in the same way. Based on three
signals captured, we establish three different way of using this information in
the classification process:

\begin{enumerate}

    \item \textit{1look}: for each object, we use signals at integration angle of
    $30^{\circ}$ only. If only one polarization is used, an object of interest can
    be identified by SRC (and implicitly, SVM). If more than one polarization is
    involved, one object will be determined by SRC-SM, as this is the best method
    based on previous experiments (see Figure~\ref{fig8}b).

    \item \textit{2look}: each object is represented by two singles captured at
    $15^{\circ}$. This multi-look classification problem can be solved by joint
    SRC (JSRC) \cite{zhang2012multi}, or the proposed relative-look SRC
    (ShiftSRC) (see Figure~\ref{fig8}c).

    \item \textit{3look}: uses all three signals to represent an object. In this
    case, the relationship between the $30^{\circ}$ signal and the first
    $15^{\circ}$ signal can be modeled by the SRC-SM, while the relationship
    between two $15^ {\circ}$ signals can be formed by either JSRC or ShiftSRC
    (see Figure~\ref{fig8}d).

\end{enumerate}

\begin{figure}[t]
    \centering
    \includegraphics[width = .49\textwidth]{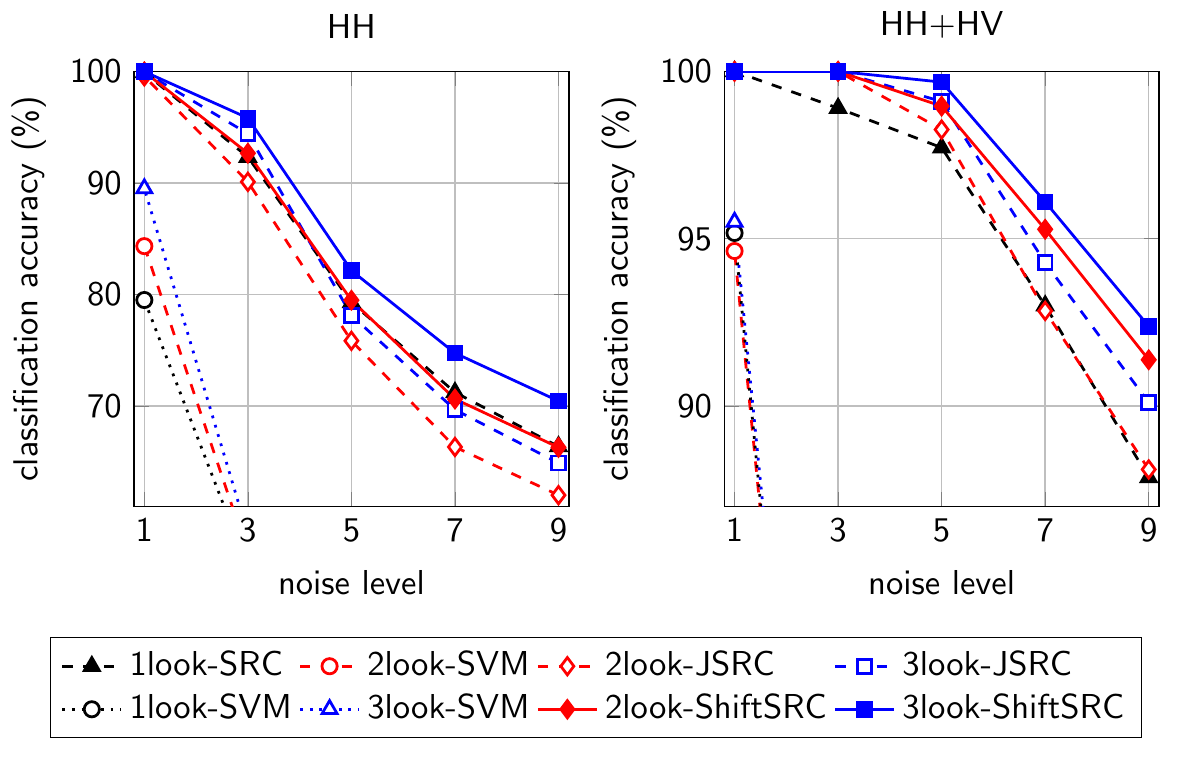}
    \caption{\small Classification accuracy (\%) of different methods on
    multiple-relative-look classification problem.}
    \label{fig9}
\end{figure}

For this experiment, we consider the \textit{all-target} scenario and two
polarization combinations, HH and HH+HV. It is worth noting that for the
HH+HV case, there will be four channels in \textit{2look} and six channels in
\textit{3look}. The results of different methods are shown in
Figure~\ref{fig9}. We can see that SVM still performs well at
the lowest noise level (1), but drastically degrades with a little more
corruption. On the other hand, SRC-based methods obtain good results
even if the noise level is large for the HH+HV combination. Of sparse representation
methods, ShiftSRC outperforms the others with the gap becoming larger for highly
corrupted signals. Interestingly, ShiftSRC at \textit{2look} provides even
better results than JSRC does at \textit{3look}. These results confirm the
advantages of the proposed relative-look sparse representation model.

\begin{figure}
    \centering
    \includegraphics[width = .49\textwidth]{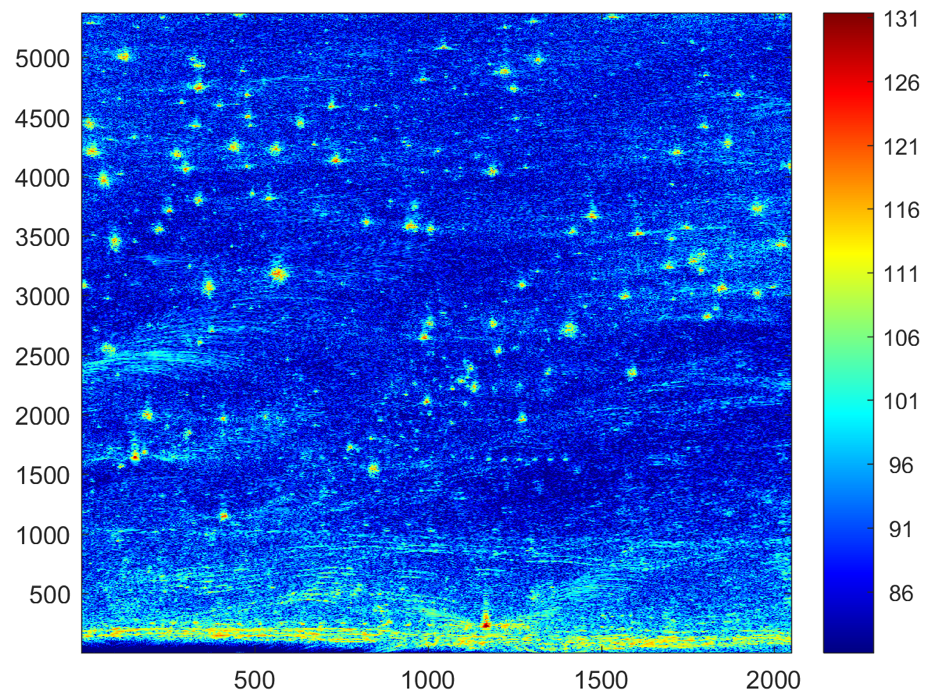}
    \caption{A VV-polarized SAR image of a minefield collected at Yuma Proving Grounds using the Army Research Laboratory UWB radar.}
    \label{fig10}
\end{figure}

\begin{figure}
    \centering
    \includegraphics[width = .47\textwidth]{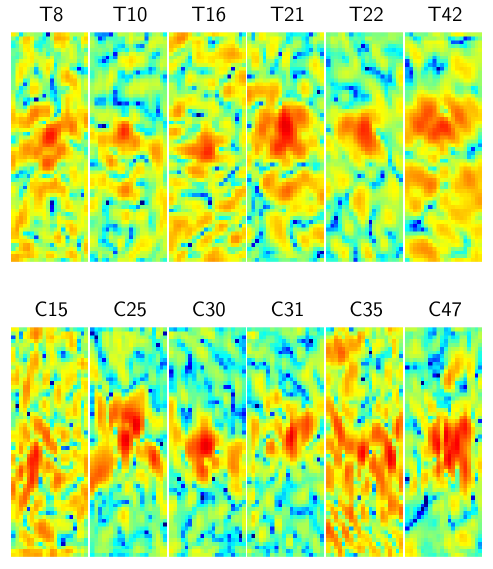}
    \caption{Visualization of real UWB SAR signals. Top: signals of targets,
    bottom: signals of confusers.}
    \label{fig11}
\end{figure}

\subsection{Overall accuracy on measured UWB SAR data} 
\label{sub:overall_accuracy_on_real_data}
In this section, the results of this technique are illustrated using the data
from the ARL UWB low-frequency SAR, which transmits
radar signals occupying the frequency spectrum that span approximately from 50
to 1150 MHz~\cite{ressler1995army}. Figure~\ref{fig10} shows a SAR image formed using data collected at
Yuma Proving Grounds (YPG)~\cite{lam1998}. The scene includes several rows of buried
mines surrounded by clutter objects such as bushes, rocks, tire tracks, etc.

\begin{figure}[t]
    \centering
    \includegraphics[width = .49\textwidth]{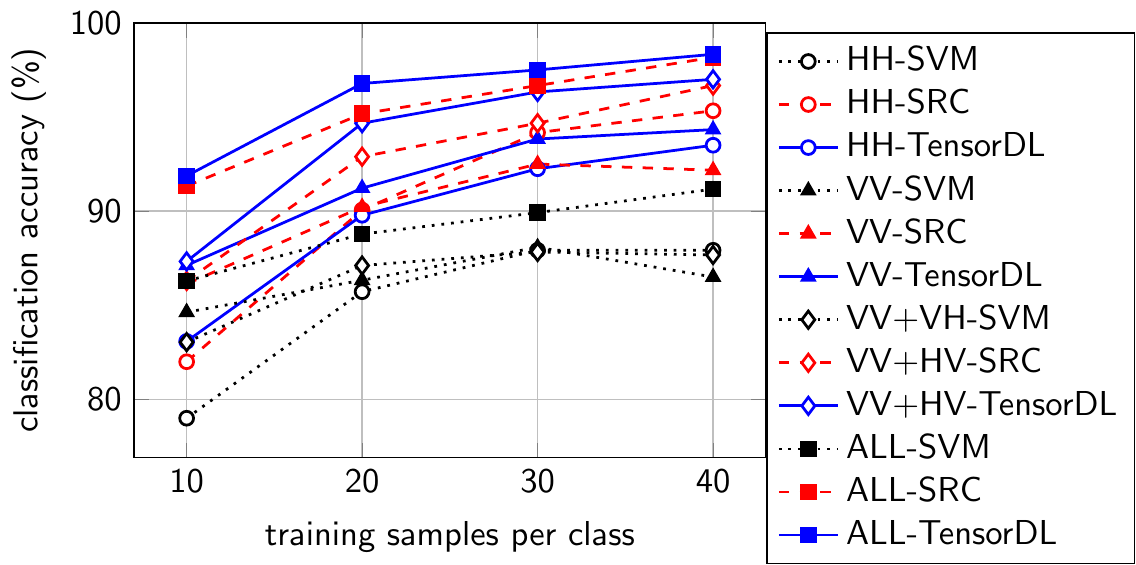}
    \caption{\small Classification accuracy on real data.}
    \label{fig12}
\end{figure}

The set contains signals of 50 targets and 50 confusers. Each signal has
resolution of $90\times 30$ and already includes noise from the ground.
Visualization of six samples in each class are shown in
Figure~\ref{fig11}. We conduct the \textit{all-target} experiment
and report results of different methods on different polarization combinations
in Figure~\ref{fig12}. For each combination, three competing methods
are considered: SVM, SRC, and TensorDL (both with the SM constraint). Since
grounds are fixed in this data set, we report the results based on size of the
training set. For each training size $N$ ($N = $10, 20, 30, or 40), we randomly
choose $N$ samples from each class for training; the rest $50 - N$ samples are
considered test samples.

The results are reported in Figure~\ref{fig12} as the average of 10
tests. In general, the tensor dictionary learning performs better than sparse
representation in all combinations except for the HH case. SVM also provides good
results but is outperformed by other {competing} methods.


\section{Conclusion} 

We have developed a novel discrimination and classification framework for 
low-frequency UWB SAR imagery using sparse representation-based methods. The
framework is applicable for either single channel or multiple channel
(polarizations, multilook) SAR imagery. The techniques are tested and the
discrimination/classification performance of targets of interest versus natural
and manmade clutter in challenging scenarios {is} measured using both rigorous
electromagnetic simulation data and real data from the ARL UWB radar. The
classification results show encouraging potential of tensor sparsity methods,
even when the test images are very noisy (buried under extremely rough ground
surfaces), targets have small RCSs, and the targets' responses have very little
detailed structures, i.e., targets are small compared to the wavelengths of the
radar signals. The SRC-SM technique and its dictionary learning version
consistently offers the best results with the combination of co- and cross-pol
data, e.g., HH+HV or ALL. In addition, the non-negativity constraints on sparse
tensor codes enhance the performance of the systems. Furthermore, we also show
the ShiftSRC model is particularly suitable for problems with
multiple-relative-look signals.

\bibliographystyle{IEEEtran}
\bibliography{TAES_2column_final}

\begin{thebibliography}{10}
\providecommand{\url}[1]{#1}
\csname url@samestyle\endcsname
\providecommand{\newblock}{\relax}
\providecommand{\bibinfo}[2]{#2}
\providecommand{\BIBentrySTDinterwordspacing}{\spaceskip=0pt\relax}
\providecommand{\BIBentryALTinterwordstretchfactor}{4}
\providecommand{\BIBentryALTinterwordspacing}{\spaceskip=\fontdimen2\font plus
\BIBentryALTinterwordstretchfactor\fontdimen3\font minus
  \fontdimen4\font\relax}
\providecommand{\BIBforeignlanguage}[2]{{%
\expandafter\ifx\csname l@#1\endcsname\relax
\typeout{** WARNING: IEEEtran.bst: No hyphenation pattern has been}%
\typeout{** loaded for the language `#1'. Using the pattern for}%
\typeout{** the default language instead.}%
\else
\language=\csname l@#1\endcsname
\fi
#2}}
\providecommand{\BIBdecl}{\relax}
\BIBdecl

\bibitem{lam1997}
L.~H. Nguyen, R.~Kapoor, and J.~Sichina, ``Detection algorithms for
  ultrawideband foliage-penetration radar,'' \emph{Proceedings of the SPIE,
  3066}, pp. 165--176, 1997.

\bibitem{lam1998}
L.~H. Nguyen, K.~Kappra, D.~Wong, R.~Kapoor, and J.~Sichina, ``Mine field
  detection algorithm utilizing data from an ultrawideband wide-area
  surveillance radar,'' \emph{SPIE International Society of Optical Engineers,
  3392}, no. 627, 1998.

\bibitem{lam2008}
L.~H. Nguyen, M.~Ressler, and J.~Sichina, ``Sensing through the wall imaging
  using the {A}rmy {R}esearch {L}ab ultra-wideband synchronous impulse
  reconstruction ({U}{W}{B} {S}{I}{R}{E}) radar,'' \emph{Proceedings of SPIE,
  6947}, no. 69470B, 2008.

\bibitem{nguyen1998ultra}
L.~Nguyen, R.~Kapoor, D.~Wong, and J.~Sichina, ``Ultra-wideband radar target
  discrimination utilizing an advanced feature set,'' \emph{SPIE, Algorithms
  for Synthetic Aperture Radar Imagery V}, 1998.

\bibitem{damarla2001detection}
T.~R. Damarla, L.~Nguyen, and K.~Ranney, ``Detection of {U}{X}{O} contaminated
  land fields using hidden {M}arkov models in the {S}{A}{R} images generated by
  airborne radar system,'' in \emph{Proceedings of SPIE}, vol. 4374, 2001,
  p.~61.

\bibitem{runkle2001multi}
P.~Runkle, L.~H. Nguyen, J.~H. McClellan, and L.~Carin, ``Multi-aspect target
  detection for sar imagery using hidden markov models,'' \emph{IEEE Trans.\ on
  Geos. and Remote Sensing}, vol.~39, no.~1, pp. 46--55, 2001.

\bibitem{anabuki2017ultrawideband}
M.~Anabuki, S.~Okumura, T.~Sato, T.~Sakamoto, K.~Saho, M.~Yoshioka, K.~Inoue,
  T.~Fukuda, and H.~Sakai, ``Ultrawideband radar imaging using adaptive array
  and doppler separation,'' \emph{IEEE Transactions on Aerospace and Electronic
  Systems}, vol.~53, no.~1, pp. 190--200, 2017.

\bibitem{sakamoto2016fast}
T.~Sakamoto, T.~Sato, P.~Aubry, and A.~Yarovoy, ``Fast imaging method for
  security systems using ultrawideband radar,'' \emph{IEEE Transactions on
  Aerospace and Electronic Systems}, vol.~52, no.~2, pp. 658--670, 2016.

\bibitem{yamaryo2018range}
A.~Yamaryo, T.~Takatori, S.~Kidera, and T.~Kirimoto, ``Range-point
  migration-based image expansion method exploiting fully polarimetric data for
  {U}{W}{B} short-range radar,'' \emph{IEEE Transactions on Geoscience and
  Remote Sensing}, vol.~56, no.~4, pp. 2170--2182, 2018.

\bibitem{kiasari2014classification}
M.~A. Kiasari, S.~Y. Na, and J.~Y. Kim, ``Classification of human postures
  using ultra-wide band radar based on neural networks,'' in \emph{IEEE
  International Conference on IT Convergence and Security (ICITCS)}, 2014, pp.
  1--4.

\bibitem{bryan2012application}
J.~Bryan, J.~Kwon, N.~Lee, and Y.~Kim, ``Application of ultra-wide band radar
  for classification of human activities,'' \emph{IET Radar, Sonar \&
  Navigation}, vol.~6, no.~3, pp. 172--179, 2012.

\bibitem{liang2018through}
X.~Liang, T.~Lv, H.~Zhang, Y.~Gao, and G.~Fang, ``Through-wall human being
  detection using {U}{W}{B} impulse radar,'' \emph{EURASIP Journal on Wireless
  Communications and Networking}, vol. 2018, no.~1, p.~46, 2018.

\bibitem{wang2015classification}
D.~Wang, L.~Chen, D.~Piscarreta, and K.~W. Tam, ``Classification and regression
  of ultra wide band signals,'' in \emph{IEEE Chinese Automation Congress
  (CAC)}, 2015, pp. 1907--1912.

\bibitem{Wright2009SRC}
J.~Wright, A.~Yang, A.~Ganesh, S.~Sastry, and Y.~Ma, ``Robust face recognition
  via sparse representation,'' \emph{IEEE Trans.\ on Pattern Analysis and
  Machine Intelligence}, vol.~31, no.~2, pp. 210--227, Feb. 2009.

\bibitem{vu2015dfdl}
T.~H. Vu, H.~S. Mousavi, V.~Monga, U.~Rao, and G.~Rao, ``{D}{F}{D}{L}:
  Discriminative feature-oriented dictionary learning for histopathological
  image classification,'' \emph{IEEE Int.\ Symposium on Biomedical Imaging},
  pp. 990--994, 2015.

\bibitem{Srinivas2013}
U.~Srinivas, H.~S. Mousavi, C.~Jeon, V.~Monga, A.~Hattel, and B.~Jayarao,
  ``{SHIRC}: A simultaneous sparsity model for histopathological image
  representation and classification,'' \emph{IEEE Int.\ Symposium on Biomedical
  Imaging}, pp. 1118--1121, Apr. 2013.

\bibitem{vu2016tmi}
T.~H. Vu, H.~S. Mousavi, V.~Monga, U.~Rao, and G.~Rao, ``Histopathological
  image classification using discriminative feature-oriented dictionary
  learning,'' \emph{IEEE Trans.\ on Medical Imaging}, vol.~35, no.~3, pp.
  738--751, March, 2016.

\bibitem{Srinivas2014SHIRC}
U.~Srinivas, H.~S. Mousavi, V.~Monga, A.~Hattel, and B.~Jayarao, ``Simultaneous
  sparsity model for histopathological image representation and
  classification,'' \emph{IEEE Trans.\ on Medical Imaging}, vol.~33, no.~5, pp.
  1163--1179, May 2014.

\bibitem{sun2015task}
X.~Sun, N.~M. Nasrabadi, and T.~D. Tran, ``Task-driven dictionary learning for
  hyperspectral image classification with structured sparsity constraints,''
  \emph{IEEE Trans.\ on Geos. and Remote Sensing}, vol.~53, no.~8, pp.
  4457--4471, 2015.

\bibitem{sun2014structured}
X.~Sun, Q.~Qu, N.~M. Nasrabadi, and T.~D. Tran, ``Structured priors for
  sparse-representation-based hyperspectral image classification,'' \emph{IEEE
  Geos. and Remote Sensing Letters}, vol.~11, no.~7, pp. 1235--1239, 2014.

\bibitem{chen2013hyperspectral}
Y.~Chen, N.~M. Nasrabadi, and T.~D. Tran, ``Hyperspectral image classification
  via kernel sparse representation,'' \emph{IEEE Trans.\ on Geos. and Remote
  Sensing}, vol.~51, no.~1, pp. 217--231, 2013.

\bibitem{zhang2012multi}
H.~Zhang, N.~M. Nasrabadi, Y.~Zhang, and T.~S. Huang, ``Multi-view automatic
  target recognition using joint sparse representation,'' \emph{IEEE Trans.\ on
  Aerospace and Electronic Systems}, vol.~48, no.~3, pp. 2481--2497, 2012.

\bibitem{mo2014adaptive}
X.~Mo, V.~Monga, R.~Bala, and Z.~Fan, ``Adaptive sparse representations for
  video anomaly detection,'' \emph{IEEE Trans.\ on Circuits and Systems for
  Video Technology}, vol.~24, no.~4, pp. 631--645, 2014.

\bibitem{vu2016icip}
T.~H. Vu and V.~Monga, ``Learning a low-rank shared dictionary for object
  classification,'' \emph{IEEE Int.\ Conference on Image Processing}, 2016.

\bibitem{Mousavi2014ICIP}
H.~S. Mousavi, U.~Srinivas, V.~Monga, Y.~Suo, M.~Dao, and T.~Tran, ``Multi-task
  image classification via collaborative, hierarchical spike-and-slab priors,''
  in \emph{IEEE Int.\ Conference on Image Processing}, 2014, pp. 4236--4240.

\bibitem{vu2016fast}
T.~H. Vu and V.~Monga, ``Fast low-rank shared dictionary learning for image
  classification,'' \emph{IEEE Trans.\ on Image Processing}, vol.~26, no.~11,
  pp. 5160--5175, Nov 2017.

\bibitem{srinivas2015structured}
U.~Srinivas, Y.~Suo, M.~Dao, V.~Monga, and T.~D. Tran, ``Structured sparse
  priors for image classification,'' \emph{IEEE Trans.\ on Image Processing},
  vol.~24, no.~6, pp. 1763--1776, 2015.

\bibitem{zhang2012joint}
H.~Zhang, Y.~Zhang, N.~M. Nasrabadi, and T.~S. Huang,
  ``Joint-structured-sparsity-based classification for multiple-measurement
  transient acoustic signals,'' \emph{IEEE Trans.\ on Systems, Man, and
  Cybernetics, Part B: Cybernetics}, vol.~42, no.~6, pp. 1586--1598, 2012.

\bibitem{dao2014structured}
M.~Dao, Y.~Suo, S.~P. Chin, and T.~D. Tran, ``Structured sparse representation
  with low-rank interference,'' in \emph{IEEE Asilomar Conference on Signals,
  Systems and Computers}, 2014, pp. 106--110.

\bibitem{dao2016collaborative}
M.~Dao, N.~H. Nguyen, N.~M. Nasrabadi, and T.~D. Tran, ``Collaborative
  multi-sensor classification via sparsity-based representation,'' \emph{IEEE
  Trans.\ on Signal Processing}, vol.~64, no.~9, pp. 2400--2415, 2016.

\bibitem{van2013design}
H.~Van~Nguyen, V.~M. Patel, N.~M. Nasrabadi, and R.~Chellappa, ``Design of
  non-linear kernel dictionaries for object recognition,'' \emph{IEEE Trans.\
  on Image Processing}, vol.~22, no.~12, pp. 5123--5135, 2013.

\bibitem{lamnguyen2013}
L.~Nguyen and C.~Le, ``Sparsity driven target discrimination for ultra-wideband
  ({U}{W}{B}) down-looking {G}round penetration radar ({G}{P}{R}),''
  \emph{Tri-Service Radar}, 2013.

\bibitem{tropp2006algorithms}
J.~A. Tropp, A.~C. Gilbert, and M.~J. Strauss, ``Algorithms for simultaneous
  sparse approximation. part i: Greedy pursuit,'' \emph{Signal Processing},
  vol.~86, no.~3, pp. 572--588, 2006.

\bibitem{beck2009fast}
A.~Beck and M.~Teboulle, ``A fast iterative shrinkage-thresholding algorithm
  for linear inverse problems,'' \emph{SIAM Journal on Imaging Sciences},
  vol.~2, no.~1, pp. 183--202, 2009.

\bibitem{vu2017dictol}
T.~H. Vu, ``{DICTOL -- A dictionary learning toolbox},''
  \url{https://github.com/tiepvupsu/DICTOL}, 2017, [Online; accessed
  15-November-2017].

\bibitem{vu2017tensor}
T.~H. Vu, L.~Nguyen, C.~Le, and V.~Monga, ``Tensor sparsity for classifying
  low-frequency ultra-wideband ({U}{W}{B}) {S}{A}{R} imagery,'' 2017, pp.
  0557--0562.

\bibitem{boyd2011distributed}
S.~Boyd, N.~Parikh, E.~Chu, B.~Peleato, and J.~Eckstein, ``Distributed
  optimization and statistical learning via the alternating direction method of
  multipliers,'' \emph{Foundations and Trends{\textregistered} in Machine
  Learning}, vol.~3, no.~1, pp. 1--122, 2011.

\bibitem{SPAMS}
``{SPA}rse {M}odeling {S}oftware,'' \url{http://spams-devel.gforge.inria.fr/},
  accessed: 2014-11-05.

\bibitem{Yu2011}
Y.~Yu, J.~Huang, S.~Zhang, C.~Restif, X.~Huang, and D.~Metaxas, ``Group
  sparsity based classification for cervigram segmentation,'' \emph{IEEE Int.\
  Symposium on Biomedical Imaging}, pp. 1425--1429, 2011.

\bibitem{yang2014sparse}
M.~Yang, L.~Zhang, X.~Feng, and D.~Zhang, ``Sparse representation based
  {F}isher discrimination dictionary learning for image classification,''
  \emph{Int.\ Journal of Computer Vision}, vol. 109, no.~3, pp. 209--232, 2014.

\bibitem{ramirez2010classification}
I.~Ramirez, P.~Sprechmann, and G.~Sapiro, ``Classification and clustering via
  dictionary learning with structured incoherence and shared features,'' in
  \emph{IEEE Int.\ Conference Computer Vision Pattern Recognition}, 2010, pp.
  3501--3508.

\bibitem{vu2017tensorsparsity}
T.~H. Vu, ``{Tensor Sparsity Toolbox},''
  \url{https://github.com/tiepvupsu/tensorsparsity}, 2017, [Online; accessed
  10-November-2017].

\bibitem{dogaru2010}
T.~Dogaru, ``A{F}{D}{T}{D} user's manual,'' \emph{ARL Technical Report,
  Adelphi, MD, ARL-TR-5145}, March 2010.

\bibitem{dogaru2007}
T.~Dogaru, L.~Nguyen, and C.~Le, ``Computer models of the human body signature
  for sensing through the wall radar applications,'' \emph{ARL, Adelphi, MD,
  ARL-TR-4290}, September 2007.

\bibitem{liao2012}
D.~Liao and T.~Dogaru, ``Full-wave characterization of rough terrain surface
  scattering for forward-looking radar applications,'' \emph{IEEE Trans.\ on
  Antenna and Propagation}, vol.~60, pp. 3853--3866, August 2012.

\bibitem{McCorkle1994}
J.~McCorkle and L.~Nguyen, ``Focusing of dispersive targets using synthetic
  aperture radar,'' \emph{U.S. Army Research Laboratory, Adelphi, MD,
  ARL-TR-305}, August 1994.

\bibitem{CC01a}
C.-C. Chang and C.-J. Lin, ``{LIBSVM}: A library for support vector machines,''
  \emph{ACM Trans.\ on Intelligent Systems and Technology}, vol.~2, pp.
  27:1--27:27, 2011, software available at
  \url{http://www.csie.ntu.edu.tw/~cjlin/libsvm}.

\bibitem{ressler1995army}
M.~Ressler, L.~Happ, L.~Nguyen, T.~Ton, and M.~Bennett, ``The army research
  laboratory ultra-wide band testbed radars,'' in \emph{IEEE International
  Radar Conference}, 1995, pp. 686--691.

\end{thebibliography}

\begin{IEEEbiography}[{\includegraphics[width=1in, clip,keepaspectratio]{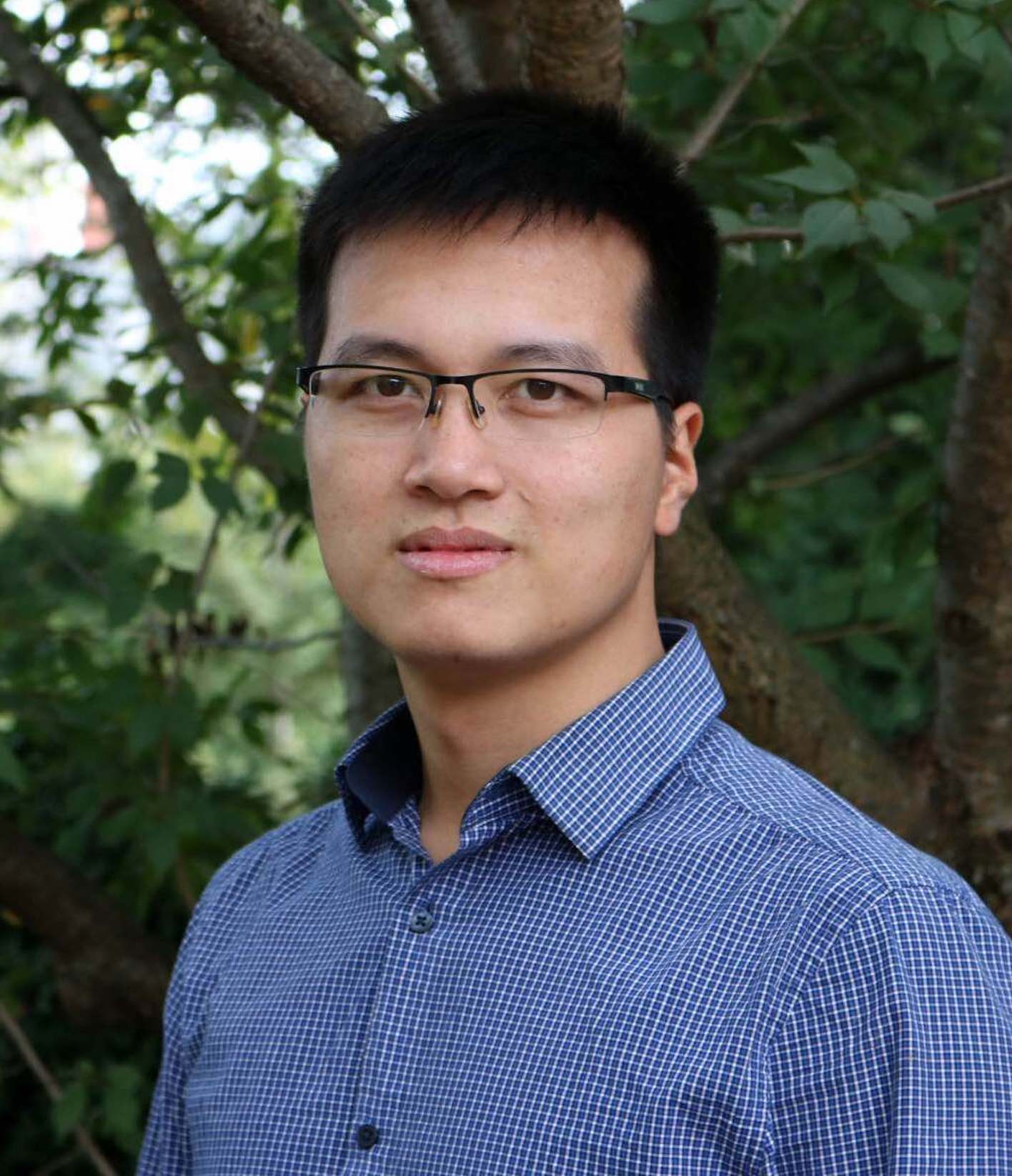}}]{Tiep Huu Vu} received the B.S. degree in Electronics and Telecommunications from Hanoi University of Science and Technology, Vietnam, in 2012. He is currently pursuing the Ph.D. degree with the information Processing and Algorithm Laboratory (iPAL), The Pennsylvania State University, University Park, PA. 

He did two Research Internships with the Army Research Lab, MD in 2016 and 2017. 
His research interests are broadly in the areas of statistical learning for signal and image analysis, computer vision and pattern recognition for image classification, segmentation, recovery and retrieval. 

\end{IEEEbiography}

\begin{IEEEbiography}[{\includegraphics[width=1in,clip,keepaspectratio]{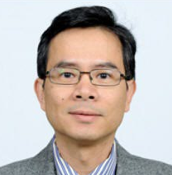}}]{Lam H. Nguyen}  received the B.S.E.E. degree from Virginia Polytechnic Institute, Blacksburg, VA, USA, the M.S.E.E. degree from George Washington University, Washington, DC, USA, and the M.S.C.S. degree from Johns Hopkins University, Baltimore, MD, USA, in 1984, 1991, and 1995, respectively.

\noindent He started his career with General Electric Company, Portsmouth, VA, in 1984. He joined Harry Diamond Laboratory, Adelphi, MD (and its predecessor Army Research Laboratory) and has worked there from 1986 to the present. Currently, he is a Signal Processing Team Leader with the U.S. Army Research Laboratory, where he has primarily engaged in the research and development of several versions of ultra-wideband (UWB) radar since the early 1990s to the present. These radar systems have been used for proof-of-concept demonstrations in many concealed target detection programs. He has been providing synthetic aperture radar (SAR) signal-processing technical consultations to industry for the developments of many state-of-the-art UWB radars. He has been developing algorithms for SAR signal and image processing. He has authored/coauthored approximately 100 conferences, journals, and technical publications. He has eleven patents in SAR system and signal processing. He has been a member of the SPIE Technical Committee on Radar Sensor Technology since 2009. He was the recipient of the U.S. Army Research and Development Achievement Awards in 2006, 2008, and 2010, the Army Research Laboratory Award for Science in 2010, and the U.S. Army Superior Civilian Performance Award in 2011.

\end{IEEEbiography}

\begin{IEEEbiography}[{\includegraphics[width=1in,clip,keepaspectratio]{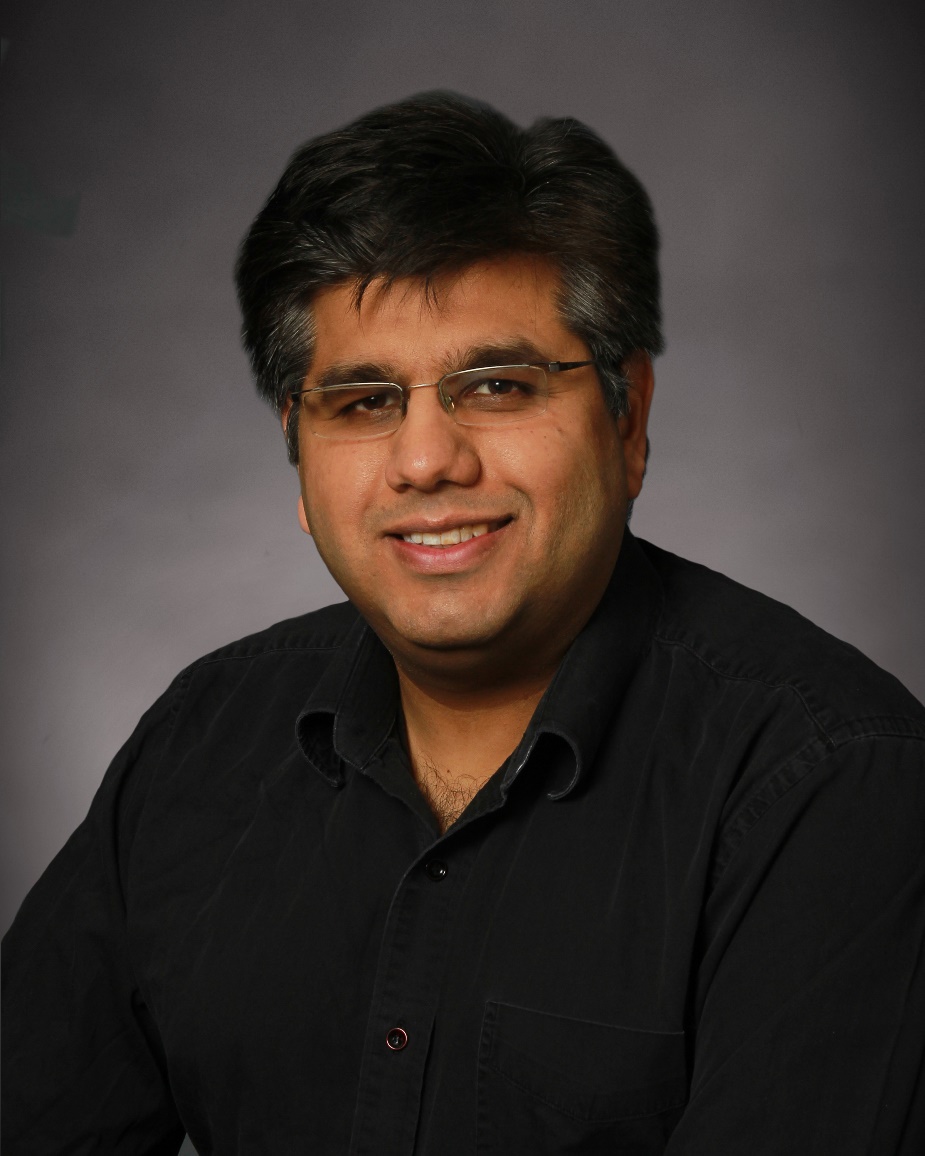}}]{Vishal Monga}
(SM  2011) is a tenured Associate Professor in the School of Electrical Engineering and Computer Science at the main campus of Pennsylvania State University in University Park, PA. He was with Xerox Research from 2005-2009. His undergraduate work was completed at the Indian Institute of Technology (IIT), Guwahati and his doctoral degree in Electrical Engineering was obtained from the University of Texas, Austin in Aug 2005.
Dr. Monga’s research interests are broadly in signal and image processing.  His research group at Penn State focuses on convex optimization approaches to image classification, robust signal (time-series) hashing, radar signal processing and computational imaging.
He currently serves as an Associate Editor for the IEEE Transactions on Image Processing, IEEE Signal Processing Letters, and the IEEE Transactions on Circuits and Systems for Video Technology. Prof. Monga is a recipient of the US National Science Foundation (NSF) CAREER award, a Monkowski Early Career award from the college of engineering at Penn State and the Joel and Ruth Spira Teaching Excellence Award.

\end{IEEEbiography}

\end{document}